\documentclass[twocolumn,a4paper]{article}
\usepackage[utf8]{inputenc}
\usepackage[T1]{fontenc}
\usepackage{amsmath,amssymb,amsfonts}
\usepackage{graphicx}
\usepackage{xcolor}
\usepackage{hyperref}
\usepackage{cite}
\usepackage{geometry}
\usepackage{abstract}
\usepackage{graphicx}
\usepackage{array}
\usepackage{subfigure}

\title{\vspace{-1.5em}\textbf{The Influence of Stable Photon Sphere Advent on Orbital Precession in moving towards the Extremality}}
\author{\textbf{Mohammad Ali S. Afshar}\textsuperscript{1,2}\thanks{m.a.s.afshar@gmail.com}, \textbf{J. Sadeghi}\textsuperscript{1}\thanks{pouriya@ipm.ir}. \\[0.5em]
\small\textsuperscript{1}Department of Theoretical Physics, Faculty of Basic Sciences, University of Mazandaran, P. O. Box 47416-95447, Babolsar, Iran \\
\small\textsuperscript{2}School of Physics, Damghan University, P. O. Box 3671641167, Damghan, Iran \\
}

\newcommand{\keywords}[1]{\par\vspace{0.3em}\noindent\textbf{Keywords:} #1}
\begin{document}
\maketitle
\thispagestyle{empty}

\begin{abstract}
In this work, we investigate the behaviour of the periapsis shift in static charged black-hole spacetimes, focusing on the interplay between extremality and the Aschenbach-like effect. We first examine the evolution of orbital dynamics as the extremal limit is approached in representative charged black-hole models. We then extend the analysis to black-hole geometries that admit a stable photon sphere outside the event horizon, where the Aschenbach-like effect has previously been identified.  Our results show that the periapsis shift remains a well-defined dynamical quantity in the extremal regime and continuously reflects changes in the underlying spacetime geometry. For black-hole solutions possessing an external stable photon sphere, the associated minimum of the effective potential produces characteristic modifications in both the angular-velocity profile and the periapsis shift. When extremality and the Aschenbach-like effect coexist, the orbital dynamics undergo a qualitative transition, giving rise to a three-region structure consisting of an inner prograde region, an intermediate retrograde region, and an outer prograde region. Within the black-hole models investigated in this work, such behaviour is absent from the other configurations considered.  These results demonstrate that the periapsis shift provides a sensitive dynamical probe of strong-field black-hole spacetimes, encoding both geometric and orbital information beyond that obtained from photon-sphere analyses alone. Our findings highlight the rich orbital structure that can emerge from the combined presence of extremality and an external stable photon sphere and provide a useful framework for future studies of orbital dynamics in modified theories of gravity.
\end{abstract}

\keywords{Periapsis Shift, Stable Photon Sphere, Aschenbach-like effect, Non rotating black holes, Extremality }

\section{Introduction}
In this section, we briefly introduce the main concepts relevant to our analysis.
\begin{center}
\textbf{Periapsis:}
\end{center}
The anomalous perihelion precession of Mercury is one of the earliest and most celebrated observational tests of general relativity, highlighting the limitations of Newtonian gravity in describing orbital motion in curved spacetime. In the Newtonian two-body problem, the orbit of a test particle around a point mass is a closed Keplerian ellipse. Within Newtonian gravity, planetary perturbations and the Sun's oblateness account for most of the observed perihelion precession. Nevertheless, these effects leave a residual advance of approximately 43 arcseconds per century unexplained \cite{1,2}. General relativity naturally accounts for this residual precession by describing gravity as a manifestation of spacetime curvature. The resulting geodesic motion predicts an additional relativistic contribution that agrees remarkably well with observations. Motivated by this classical result, it is reasonable to expect even more pronounced effects in the vicinity of ultra-compact gravitational structures. The study of periapsis shift has therefore been naturally extended to black-hole spacetimes and has been explored for a broad class of gravitational backgrounds \cite{3,4,5,6,7,8,9,10,11}.
\begin{center}
\textbf{Extremality}
\end{center}
It is well established that evaporation through radiation constitutes an intrinsic component of a black hole’s evolutionary trajectory. For charged black holes, under typical initial conditions, the charge-to-mass ratio $(Q/M)$ is invariably significantly less than one. Consequently, mass loss via radiative processes inevitably drives the system toward a state in which the charge-to-mass ratio exceeds unity, thereby reaching the extremal limit.\\An extremal black hole corresponds to the limiting configuration in which the inner and outer horizons coincide and the Hawking temperature vanishes \cite{12,13,14,15}. Because this limit represents the boundary between ordinary black holes and horizonless configurations, understanding the behaviour of physical observables in its vicinity is of particular theoretical interest.
\begin{center}
\textbf{Aschenbach-Like Effect}
\end{center}
The Aschenbach effect is a relativistic phenomenon associated with orbital motion in the strong gravitational field of rapidly rotating black holes. It was originally identified in studies of test-particle motion around near-extremal Kerr black holes. Aschenbach showed that the orbital velocity of test particles may exhibit a non-monotonic radial profile close to the event horizon. \cite{22,23}. This unexpected behaviour has attracted considerable attention because it departs from the monotonic velocity profile predicted by standard Keplerian motion. For Kerr black holes, this behaviour originates from the interplay between strong gravitational fields and frame dragging. The effect has since been investigated in several rotating black-hole spacetimes, including asymptotically flat, AdS and dS geometries \cite{23}. This naturally raises the question of whether a similar velocity profile can also arise in static black-hole spacetimes, where frame dragging is absent. Although static black holes do not possess frame dragging, recent studies have shown that non-monotonic orbital velocity profiles may still develop when the spacetime admits a stable gravitational potential minimum outside the event horizon. Such minima are typically associated with the presence of a stable photon sphere \cite{27,28,29}.\\Although both rotating and static black holes may exhibit similar non-monotonic velocity profiles, the underlying physical mechanisms are fundamentally different. In Kerr black holes, the effect originates from frame dragging, whereas in static black holes it is driven by the existence of a stable gravitational potential minimum outside the event horizon \cite{27,28,29}. Consequently, since the underlying physical origin differs from that of the Kerr case, we refer to this phenomenon as the \emph{Aschenbach-like effect}

\begin{center}
\textbf{Article Objectives}
\end{center}
Our study is motivated by the observation that several black-hole models preserve their essential geometric and dynamical properties even as they approach the extremal limit \cite{12,13,14,15,16,17,18,19,20,21}. Previous investigations have shown that, under suitable conditions and toward extremality form of black holes:\\
\begin{itemize}
    \item the spacetime geometry remains regular, the event horizon persists, and the Hawking temperature approaches zero \cite{12,13,30}.
    \item the unstable photon sphere remains outside the event horizon even in the extremal regime \cite{12,13,31}.
    \item the Aschenbach-like effect, whenever present, survives the approach to extremality without losing its characteristic non-monotonic behaviour \cite{32}.
\end{itemize}
Together, these results suggest that the qualitative properties of the black-hole spacetime remain largely unchanged as extremality is approached. Motivated by the considerations outlined above, this work is designed to address three related questions.\\
In the present work, we investigate whether another fundamental orbital observable, namely the periapsis shift, exhibits a similarly robust behaviour as the extremal limit is approached.\\
Our second objective is to examine the influence of the Aschenbach-like effect on the periapsis shift. To this end, we consider the Born--Infeld massive gravity model, which admits a stable gravitational potential minimum outside the event horizon. We investigate how this additional geometric structure modifies orbital precession relative to more conventional black-hole spacetimes.\\
Finally, we investigate black-hole models that simultaneously admit both extremality and the Aschenbach-like effect. This allows us to isolate the combined influence of these two features on the periapsis shift and to determine whether their interplay gives rise to qualitatively new orbital behaviour.
\section{Methodology}
We begin by introducing the general spacetime geometry used throughout this work. We consider a four-dimensional static and spherically symmetric spacetime. Owing to spherical symmetry, the motion can be restricted to the equatorial plane $(\theta=\pi/2)$ without loss of generality. In such a spacetime, the most general form of the line element is given by:
\begin{equation}\label{(1)}
\begin{split}
&\mathit{ds}^{2}=-g_{\mathit{tt}} \mathit{dt}^{2} +g_{\mathit{rr}} \mathit{dr}^{2} +g_{\theta \theta} d\theta^{2} +g_{\phi \phi} d\phi^{2}=\\ &-e^{\nu(r)}dt^{2}+e^{\lambda(r)}dr^{2}+r^{2}(d\theta^{2}+\sin^{2}\theta d\phi^{2}).\\
\end{split} 
\end{equation}
Throughout this paper, we use geometrized units with $G=c=\hbar=1$.

\subsection{Null and Timelike Geodesics}
The effective potential derived from the spacetime metric governs the geodesic structure and orbital dynamics around a black hole. Depending on the spacetime geometry, both open and bound trajectories may exist. Circular null and timelike geodesics correspond to photon spheres and timelike circular orbits (TCOs), respectively. Their existence and stability depend on the properties of the underlying spacetime. These properties are determined from the extrema of the effective potential and the sign of its second radial derivative. Local minima correspond to stable orbits, whereas local maxima represent unstable ones. In subsequent sections, we will elaborate on the precise mathematical equations employed to compute and analyze these orbital configurations.

\begin{center}
\textbf{Timelike Geodesics:}
\end{center}
According to the metric function Eq. (\ref{(1)}) the Lagrangian for a neutral test particle is:
\begin{equation}
\mathcal{L}=\frac{1}{2}\left[-e^{\nu}\dot{t}^{2}+e^{\lambda}\dot{r}^{2}+r^{2}(\dot{\theta}^{2}+\sin^{2}\theta \dot{\phi}^{2})\right].
\label{(2)}
\end{equation}
Restricting the motion to the equatorial plane $(\theta=\pi/2)$, the cyclic coordinates $t$ and $\phi$ give rise to the conserved quantities:
\begin{equation}
\mathbb{E}=e^{\nu}\dot{t}=constant,\quad \quad \quad \mathbb{L}={r^{2}}\dot{\phi}=constant,
\label{(3)}
\end{equation}
where $\mathbb{E}$ and $\mathbb{L}$ is energy and angular momentum respectively. One can find that the equation of motion in terms of $\mathbb{E}$ and $\mathbb{L}$ can be as follows \cite{3,4}:
\begin{equation}
\dot{r}^{2}+e^{-\lambda }\frac{\mathbb{L}^{2}}{r^{2}}=e^{-\lambda }\left(\frac{\mathbb{E}^{2}
}{c^{2}}e^{-\nu }-1\right).  
\label{(4)}
\end{equation}
Comparing Eq.~(\ref{(4)}) with the standard form $\frac{1}{2}\dot{r}^{2}+V(r)=0$, the effective potential becomes:
\begin{eqnarray}
 \mathbb{V}(r)=\frac{1}{2}e^{-\lambda}\left[\left(1+\frac{\mathbb{L}^{2}}{r^{2}}\right)-e^{-\nu}\mathbb{E}^{2}\right]. 
\label{(5)}
\end{eqnarray}
TCOs satisfy the conditions $\mathbb{V}=\mathbb{V}'=0$, which determine the conserved quantities and orbital radius. Their stability is governed by $\mathbb{V}''$: positive, negative and zero values correspond to stable, unstable and marginally stable circular orbits, respectively. Before ending this subsection, it is worth introducing another useful quantity that is very applicable to our study and can well separate areas with physical meaning from purely mathematical solutions.
\begin{eqnarray}
\beta=2-r \nu'. 
\label{(6)}
\end{eqnarray}
This quantity naturally appears in the expressions for the conserved energy and angular momentum. The physically admissible region is characterized by $\beta>0$, whereas $\beta<0$ leads to imaginary values of these conserved quantities.

\begin{center}
\textbf{Photon Sphere:}
\end{center}
Rather than employing the conventional effective-potential approach, we identify photon spheres using the recently developed topological method \cite{24,25,26}. Topological methods have recently emerged as powerful tools for investigating black-hole properties, including thermodynamics, phase transitions and photon-sphere structures \cite{32.1,32.2,33,33.1,34,34.1,35,35.1,36,37,37.1,38,38.1,39,39.1,40,40.1,41,41.1,41.2,42,42.1,43}. The method introduces a scalar potential constructed solely from the metric, which is then mapped onto a two-dimensional vector field. Photon spheres are identified with the zeros of this field, while their topological charges are determined from the winding numbers around these zeros \cite{24,25,26}. Since the formalism has been presented in detail elsewhere \cite{24,25,26}, only the essential relations are summarized below. Given the metric function Eq. (\ref{(1)}), we consider the following scalar potential \cite{24,25,26}:
\begin{equation}\label{(7)}
\begin{split}
\mathbb{H}=\sqrt{\frac{-g_{tt}}{g_{\phi\phi}}}.
\end{split}
\end{equation}
Now, based on this scalar potential function, we construct the two-dimensional vector field $\Psi=(\Psi^r,\Psi^\Theta)$.
\begin{equation}\label{(8)}
\begin{split}
&\Psi^r=\frac{\partial_rH}{\sqrt{g_{rr}}},\\
&\Psi^\Theta=\frac{\partial_\theta H}{\sqrt{g_{\theta\theta}}}.
\end{split}
\end{equation}
Accordingly, the normalized vector can be defined as,
\begin{equation}\label{(9)}
\mathbf{n}^o=\frac{\Psi^o}{||\Psi||},
\end{equation}
where $o=1,2$  and  $(\Psi^1=\Psi^r)$ , $(\Psi^2=\Psi^\Theta)$. Each zero of the vector field corresponds to a photon sphere. Stable and unstable photon spheres are associated with local minima and maxima of $\mathbb{H}$ and carry topological charges of $+1$ and $-1$, respectively. The total topological charge is given by the algebraic sum of all individual charges.
In many black-hole solutions, the total topological charge is $-1$, reflecting the presence of an unstable photon sphere outside the event horizon. In contrast, horizonless ultra-compact objects often possess vanishing total charge owing to the appearance of stable photon spheres.\\
\underline{This classification is not universal.} As shown in our previous work \cite{27} and in independent studies \cite{29}, several black-hole solutions, including models in massive gravity, Gauss--Bonnet gravity and nonlinear electrodynamics, admit stable photon spheres outside the event horizon while remaining genuine black holes. Consequently, the total topological charge may become $0$ or even $+1$ without implying a naked singularity. Such exceptional cases are examined in Secs.III and IV.

\subsection{The Periapsis Shift of a Quasi-Circular Orbit}
The radial equation of motion follows directly from the Euler--Lagrange equations and can be expressed in terms of the effective potential as \cite{3,4,11}:
\begin{equation}
\ddot{r}+\mathbb{V}'(r)=0.
\label{(10)}
\end{equation}
Given the conditions governing the effective potential for having a circular orbit, namely $\mathbb{V}=\mathbb{V}'=0$, the radius of the circular orbit can be calculated at $ r=r_{cir}$ for the model under study. Now, suppose the circular orbit is perturbed by a very tiny $\delta r$ as $r = r_{cir} + \delta r$. Expanding Eq.~(\ref{(10)}) to first order in $\delta r$ yields:
\begin{equation}
\ddot{\delta r} + \mathbb{V}^{\prime \prime}(r_{cir}) \delta r = 0. 
\label{(11)}
\end{equation}
Given the oscillatory structure of Eq. (\ref{(11)}), the radial frequency of such a motion can be written in the following form:
\begin{equation}
\omega_{r}=\sqrt{\mathbb{V}''},
\label{(12)}
\end{equation}
Also, according to Eq. (\ref{(3)}), the orbital angular velocity can be obtained as follows:
\begin{equation}
\omega_{\phi}=\sqrt{\frac{\nu'}{r(2-r\nu')}} 
\label{(13)}
\end{equation}
Combining Eqs.~(\ref{(12)}) and (\ref{(13)}), the periapsis shift is obtained as \cite{3,4,11}:
\begin{equation}
\Delta\Phi_{p}=2\pi \left(\frac{1}{\sqrt{A}}-1\right)=2\pi \left(\frac{1}{1-\sigma}-1\right),
\label{(14)}
\end{equation}
where $A=(\omega_r/\omega_\phi)^2$, and $\sigma$ denotes the periapsis advance parameter.\\The parameter $A$ determines the direction of orbital precession. For $0<A<1$, one has $\omega_\phi>\omega_r$, resulting in a positive periapsis shift ($\Delta\Phi_p>0$) corresponding to prograde precession. Conversely, for $A>1$, the radial frequency exceeds the orbital frequency ($\omega_r>\omega_\phi$), leading to a negative periapsis shift ($\Delta\Phi_p<0$) and hence retrograde precession.
\section{The Periapsis Shift and Extremality}
\begin{figure}
\begin{center}
\subfigure[]{
\includegraphics[width=\linewidth]{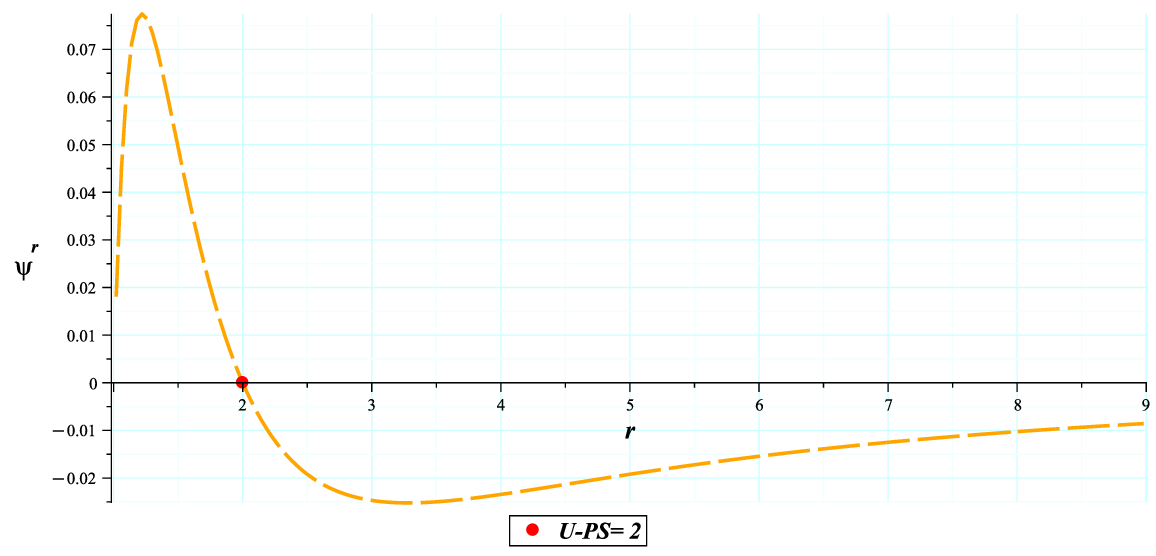}
\label{2a}}
\subfigure[]{
\includegraphics[width=\linewidth]{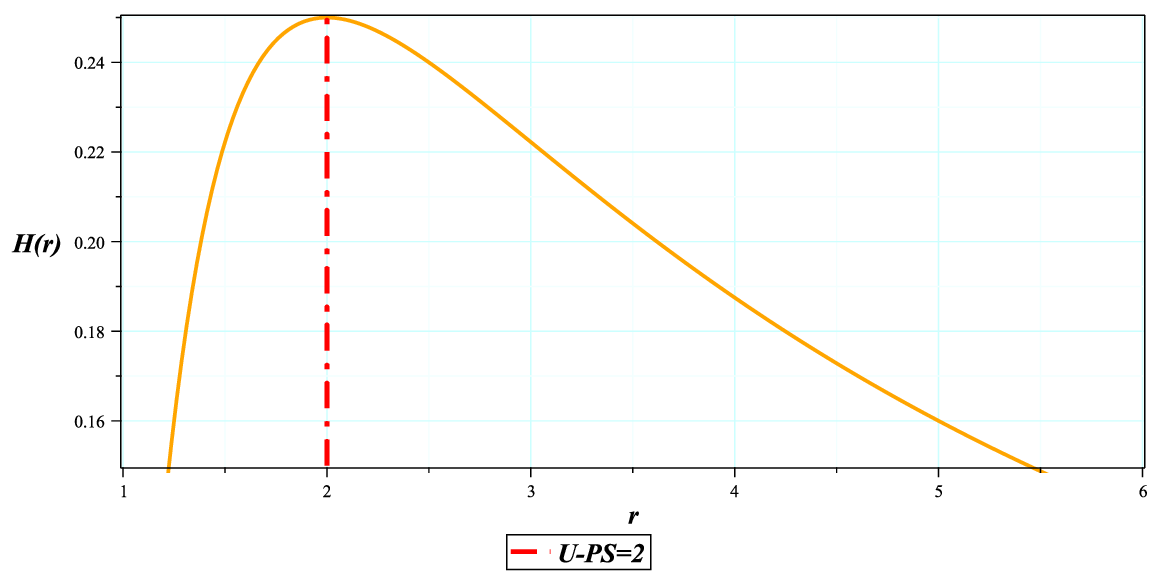}
\label{2b}}
\caption{\small{(2a): The unstable photon spheres(U-PS) with respect to $ M = 1, \gamma = 2,q = 2.718281 $ , (2b): the topological potential H(r) for ModMax Black Hole }}
\label{2}
\end{center}
\end{figure}

The RN black hole possesses a well-defined extremal configuration at $Q=M$, where the inner and outer horizons merge into a degenerate event horizon. For $Q>M$ , however, no event horizon exists and the spacetime becomes horizonless. Although the RN solution provides the canonical example of an extremal charged black hole, its electromagnetic sector is described by standard Maxwell electrodynamics. It is therefore natural to ask how modifications of the electromagnetic field influence the orbital properties of black holes that still retain their black hole properties in the form $Q>M$. For this purpose, we consider the ModMax model, which deforms the Maxwell sector while preserving the Einstein gravitational dynamics. In this theory, the effective charge is replaced by $Q^{2}e^{-\gamma}$, where $\gamma$ denotes the deformation parameter. In the limit $\gamma\rightarrow0$, the standard RN solution is recovered. In this section, we investigate the ModMax black hole in both asymptotically flat and anti-de Sitter spacetimes. Particular attention is devoted to the extremal limit, where we examine the persistence of the event horizon, the vanishing of the Hawking temperature, and the behaviour of the periapsis shift. We also determine whether the corresponding photon-sphere structure remains consistent with the geometric properties identified in previous studies. 
\begin{center}
\textbf{4D ModMax Black Hole:}
\end{center}
The spherically-symmetric static metric line element of such a spacetime is given by \cite{44}:
\begin{equation}\label{(15)}
e^{\nu(r)}=e^{-\lambda(r)}=1-\frac{2 M}{r}+\frac{q^{2} {\mathrm e}^{-\gamma}}{r^{2}}
\end{equation}
where $M$ and $q$ are the mass and charge of the black hole and $\gamma$ are known as ModMax’s dimensionless parameter.
Our studies show that the structure, for the chosen parameters, clearly has a Cauchy and event horizon up to $q = 2.71828182846$ . In this limit, the two horizons merge and the horizon temperature tends to zero, Fig. (\ref{2}).
\begin{figure}
\begin{center}
\subfigure[]{
\includegraphics[width=\linewidth]{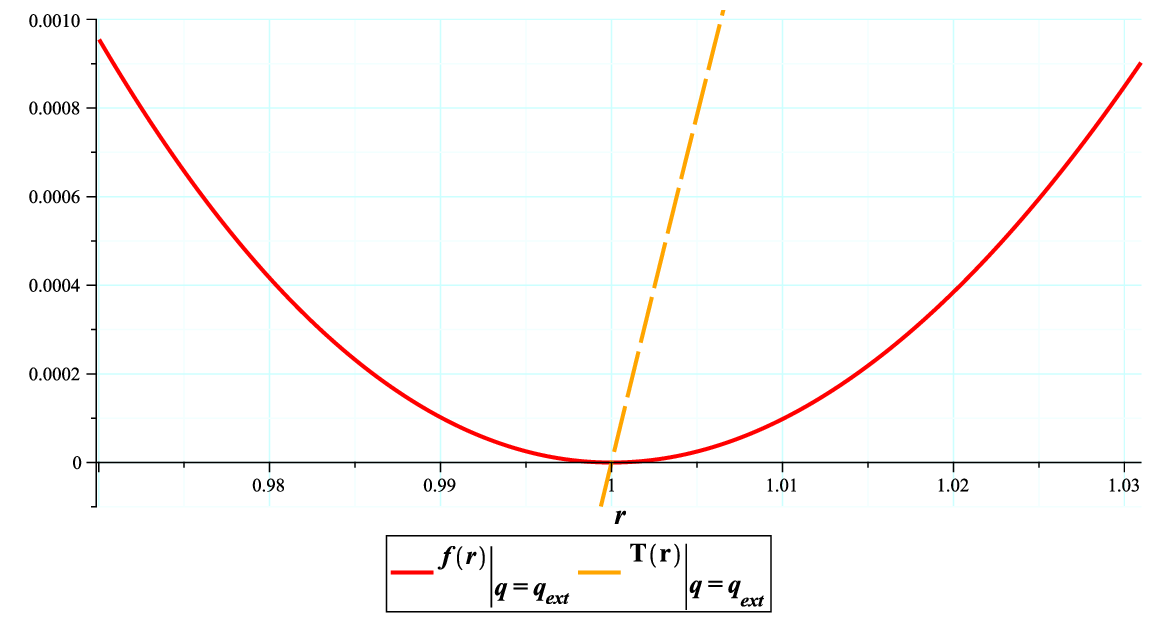}
 }
\caption{\small{ Metric function VS temperature at the extreme limit }}
\label{1}
\end{center}
\end{figure}
In this case, with using equation (\ref{(7)}), (\ref{(8)}) for $H$ and $\Psi$, we have:
\begin{equation}\label{(17)}
H =\frac{\sqrt{1-\frac{2 M}{r}+\frac{q^{2} {\mathrm e}^{-\gamma}}{r^{2}}}}{\sin \! \left(\theta \right) r},
\end{equation}
\begin{equation}\label{(18)}
\Psi^{r}=\frac{\left(-2 q^{2} {\mathrm e}^{-\gamma}+3 m r -r^{2}\right) \csc \! \left(\theta \right)}{r^{4}}.
\end{equation}
In Fig. (\ref{2a}), $\Psi^{r}$ has only one root outside the horizon, which appears as a maximum in the potential diagram, Fig. (\ref{2b}). Therefore, our photon sphere is an unstable photon sphere (U-PS), and accordingly, the structure has standard black hole behavior. So far, as we have seen, it seems that the model under study has retained all its black hole properties in the extreme limit.

\begin{figure}
\begin{center}
\subfigure[]{
\includegraphics[width=\linewidth]{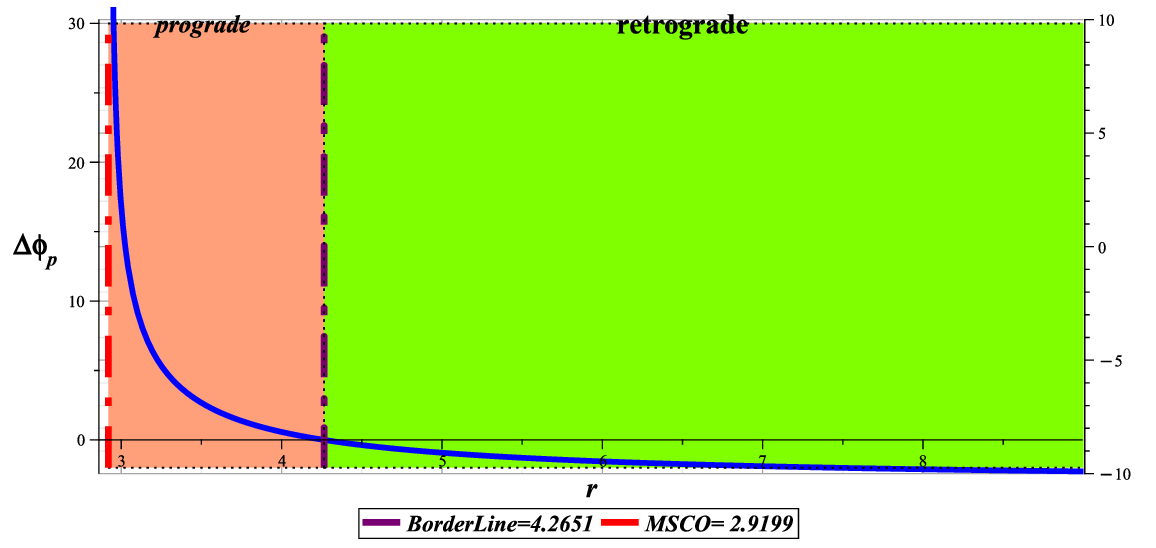}
\label{8a}}
\caption{\small{  $\Delta \Phi_{p}$ diagram with respect to $ M = 1, \gamma = 2 ,\Lambda = -1,q = 2.48019 $  for AdS ModMax Black Hole}}
\label{8}
\end{center}
\end{figure}

\begin{center}
\textbf{The Periapsis Shift}
\end{center}
To gain a deeper understanding, it is first useful to examine the geodetic behavior of the model. According to Eq. (\ref{(5)}) and Eq. (\ref{(6)}) for this model, we have:
\begin{equation}\label{(19)}
\mathbb{V}''=\frac{-4 {\mathrm e}^{-2 \gamma} q^{4}+9 q^{2} {\mathrm e}^{-\gamma} r M +r^{3} M -6 r^{2} M^{2}}{\left(2 q^{2} {\mathrm e}^{-\gamma}+r^{2}-3 M r \right) r^{4}}
\end{equation}
\begin{equation}\label{(20)}
\beta =\frac{4 q^{2} {\mathrm e}^{-\gamma}+2 r^{2}-6 M r}{q^{2} {\mathrm e}^{-\gamma}+r^{2}-2 M r}
\end{equation}
\begin{figure}
\begin{center}
\subfigure[]{
\includegraphics[width=\linewidth]{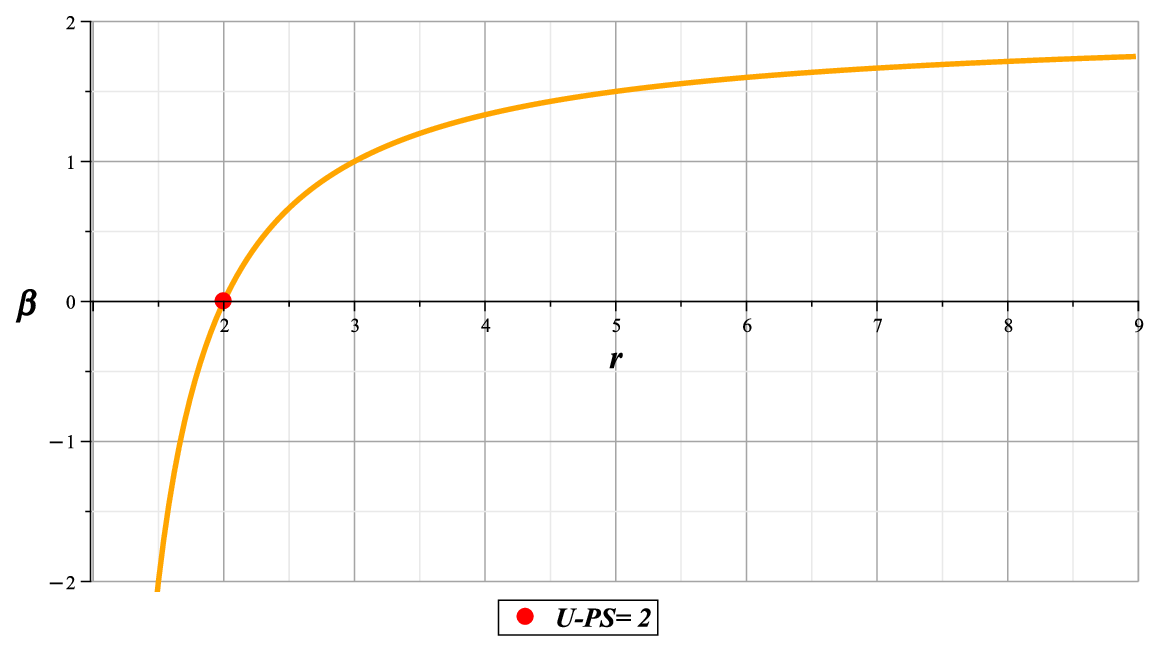}
\label{3a}}
\subfigure[]{
\includegraphics[width=\linewidth]{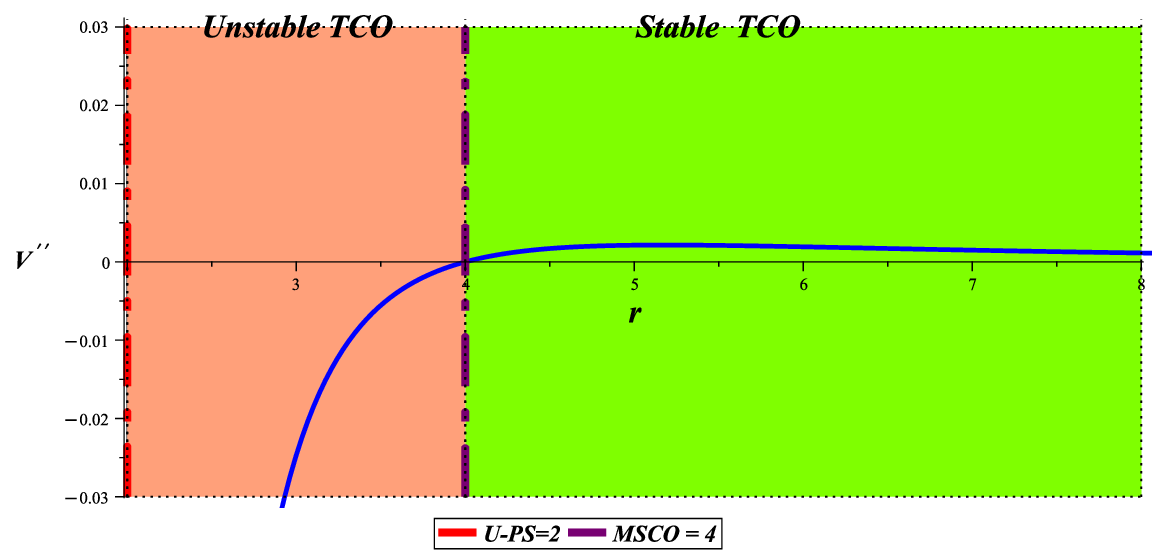}
\label{3b}}
\caption{\small{ (4a): $\beta$ diagram  (6b): MSCO localization and space classification with respect to $ M = 1, \gamma = 2,q = 2.71828 $ for ModMax Black Hole}}
\label{3}
\end{center}
\end{figure}
As can be seen in Fig. (\ref{3a}), the physically meaningful region for massive particles is from the unstable photon sphere onwards. If we consider the MSCO as a stable circular orbit of the smallest radius that is continuously connected to a set of stable TCOs, the black hole-like behavior of the model can be observed in Fig. (\ref{3b}).
Now, due to Eq. (\ref{(14)}) one can write the $\Delta\Phi_{p}$ and A for this model as:
\begin{equation}\label{(21)}
\Delta \Phi_{p} =2 \left(\frac{\sqrt{-\frac{r^{2} \left(q^{2} {\mathrm e}^{-\gamma}-M r \right)}{2 q^{2} {\mathrm e}^{-\gamma}+r^{2}-3 M r}}}{r^{2} \sqrt{\frac{-4 {\mathrm e}^{-2 \gamma} q^{4}+9 q^{2} {\mathrm e}^{-\gamma} r M+r^{3} M-6 r^{2} M^{2}}{\left(2 q^{2} {\mathrm e}^{-\gamma}+r^{2}-3 M r \right) r^{4}}}}-1\right) \pi 
\end{equation}
\begin{equation}\label{(22)}
A =-\frac{-4 {\mathrm e}^{-2 \gamma} q^{4}+9 q^{2} {\mathrm e}^{-\gamma} r M +r^{3} M-6 r^{2} M^{2}}{r^{2} \left(q^{2} {\mathrm e}^{-\gamma}-M r \right)}. 
\end{equation}
\begin{figure}
\begin{center}
\subfigure[]{
\includegraphics[width=\linewidth]{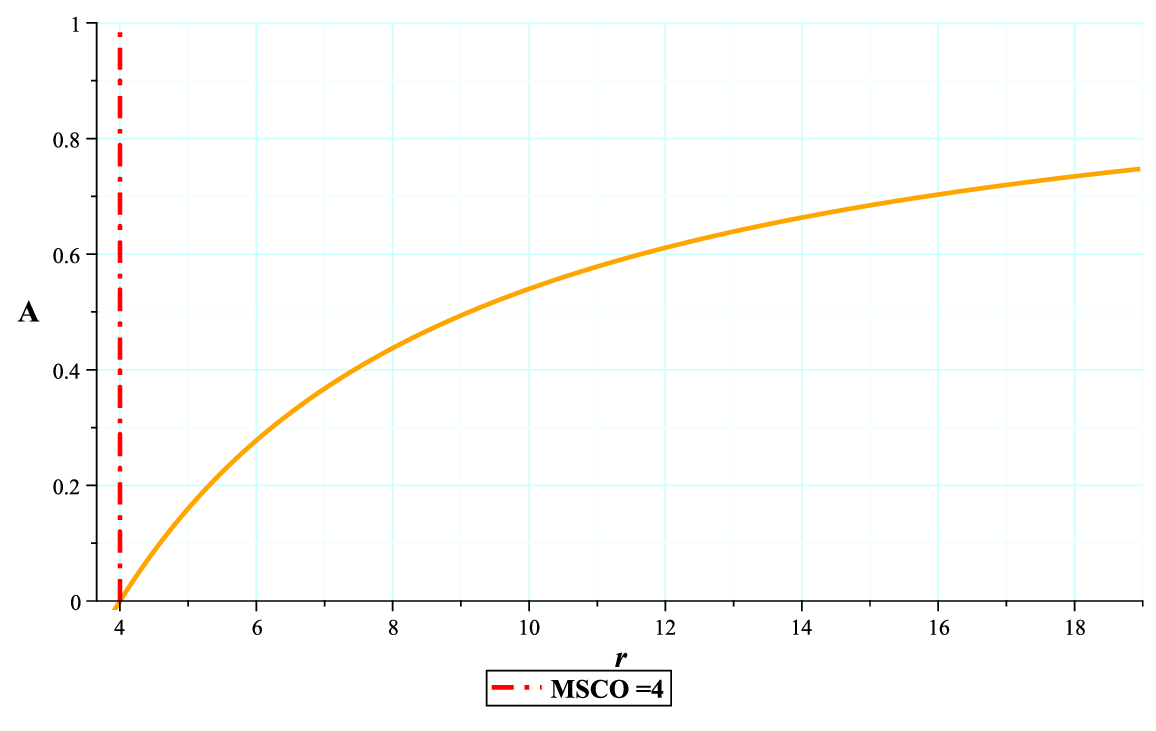}
\label{4a}}
\subfigure[]{
\includegraphics[width=\linewidth]{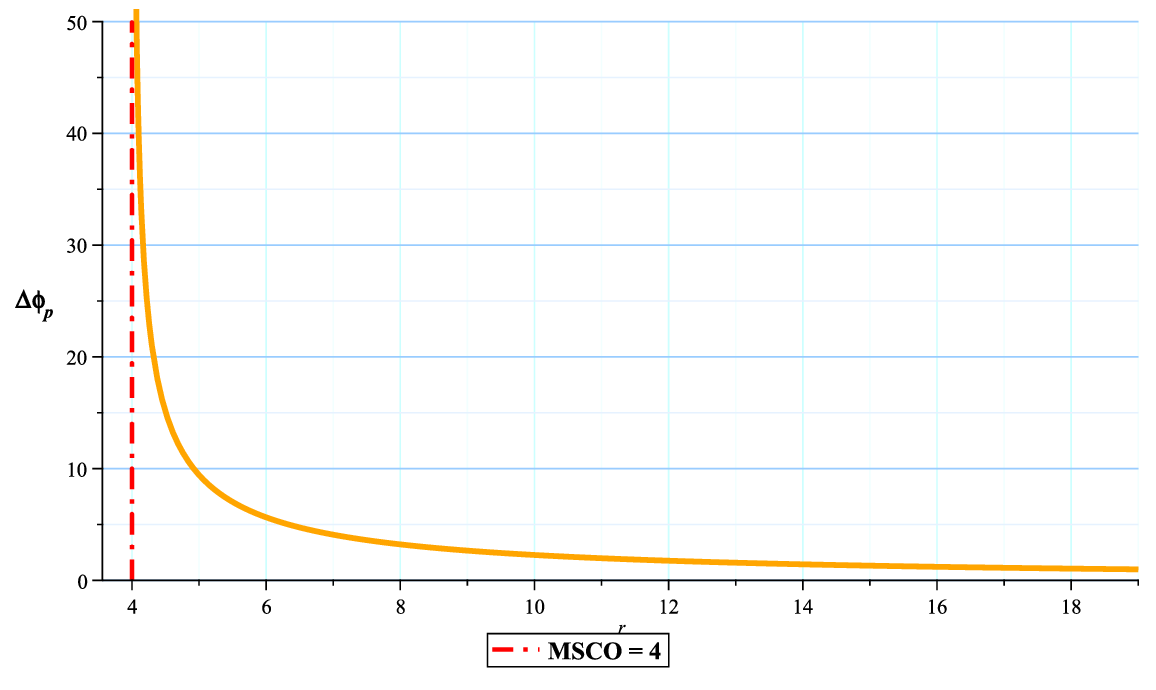}
\label{4b}}
\caption{\small{ (5a): "$A$" diagram  (5b): $\Delta \Phi_{p}$ diagram with respect to $ M = 1, \gamma = 2,q = 2.71828 $ for ModMax Black Hole}}
\label{4}
\end{center}
\end{figure}
 Fig. (\ref{4a}) reveals that the  periapsis shift initiates precisely at the onset of stable orbit formation. In this model, the parameter "A" never exceeds unity, a characteristic corroborated by the absence of negative regions in the $\Delta \Phi_{p}$ curve of Fig. (\ref{4b}). The behavior of the periapsis shift in Fig. (\ref{4}) aligns with classical expectations in sub exterm form of black hole, exhibiting a monotonic decrease as one moves away from the strong gravitational field. Most notably, the periapsis shift remains consistently prograde across the orbital configuration, with no occurrence of retrograde.\\
\begin{figure}
 \begin{center}
\subfigure[]{
\includegraphics[width=\linewidth]{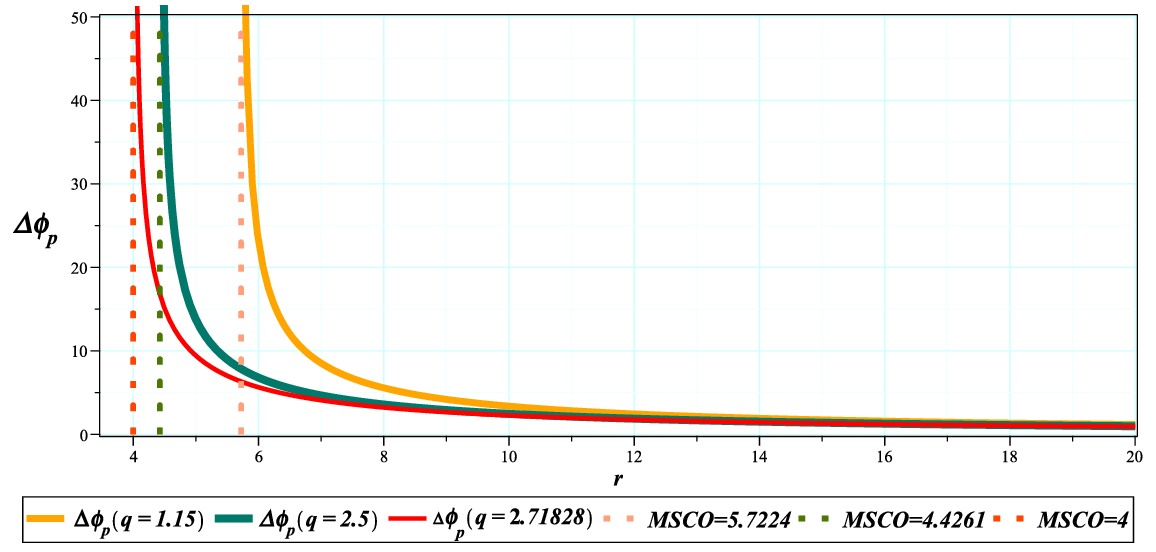}
\label{5a}}
\subfigure[]{
\includegraphics[width=\linewidth]{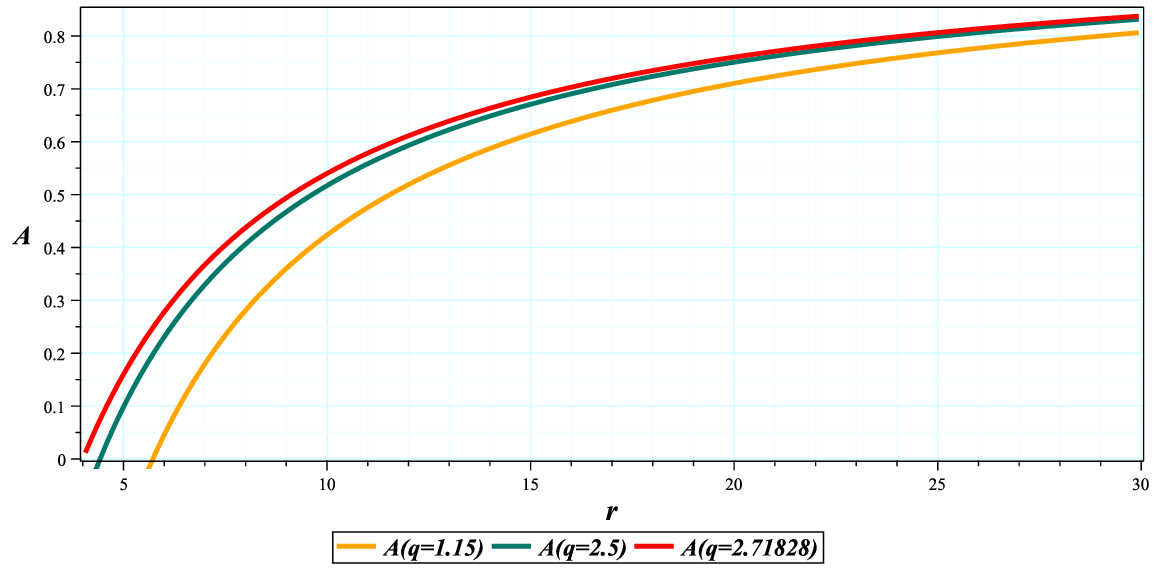}
\label{5b}}
\subfigure[]{
\includegraphics[width=\linewidth]{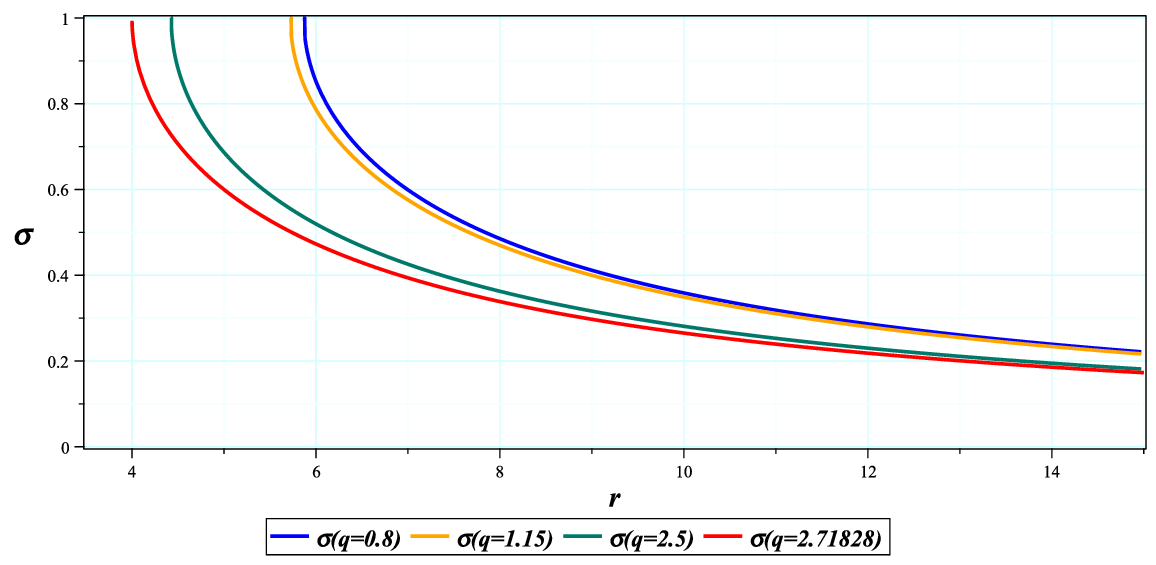}
\label{5c}}
\caption{\small{ (5a): $\Delta \Phi_{p}$ diagram  (5b): "$A$" diagram  (5c): The $\sigma$ diagram with respect to $ M = 1, \gamma = 2 $ with different q's for ModMax Black Hole}}
\label{5}
\end{center}
\end{figure}
As illustrated in  Fig. (\ref{5}), the orbital behavior of massive particles around this model continues to adhere to the canonical black hole pattern, even as the system approaches and reaches the extremal limit. So that the periapsis shift $\Delta \Phi_p$ never attains negative values and that all orbital motion remains purely prograde. 
\begin{center}
\textbf{4D AdS ModMax Black Hole:}
\end{center}
To investigate the influence of spacetime curvature, we extend our analysis from asymptotically flat spacetime to the AdS case \cite{45}. Since our primary objective is to examine how the periapsis shift evolves as the black hole approaches extremality, we do not repeat the standard analysis of the extremal configuration in the main text. Instead, the verification of the extremal limit—including the coincidence of the horizons, the vanishing of the Hawking temperature, and the persistence of an unstable photon sphere outside the event horizon—is summarized in Appendix A. Here, we focus directly on the evolution of the periapsis shift with the model parameters.  The applied equations Eq. (\ref{(5)}), Eq. (\ref{(6)}) and Eq. (\ref{(14)}) for such a model will be as follows:
\begin{equation}\label{(24)}
\beta =\frac{12 q^{2} {\mathrm e}^{-\gamma}+6 r \left(r -3 M \right)}{3 q^{2} {\mathrm e}^{-\gamma}-r \left(\Lambda  r^{3}+6 M -3 r \right)},
\end{equation}
\begin{equation}\label{(25)}
\begin{split}
&\mathbb{V}''=\frac{1}{6 \left(q^{2} {\mathrm e}^{-\gamma}+\frac{r \left(r -3 M \right)}{2}\right) r^{4}}\times\\ &[-12 q^{2} \left(\Lambda  r^{3}-\frac{9 M}{4}\right) r {\mathrm e}^{-\gamma}-4 \Lambda  r^{6}+\\ & 15 \Lambda  r^{5} M -12 {\mathrm e}^{-2 \gamma} q^{4}+3 r^{3} M -18 r^{2} M^{2}],\\
\end{split}
\end{equation}
\begin{equation*}\label{(26)}
\begin{split}
&\zeta =-12 q^{2} \left(\Lambda  r^{3}-\frac{9 M}{4}\right) r {\mathrm e}^{-\gamma}-4 \Lambda  r^{6}\\&+15 \Lambda  r^{5} M -12 {\mathrm e}^{-2 \gamma} q^{4}+3 r^{3} M -18 r^{2} M^{2},\\
\end{split} 
\end{equation*}
\begin{equation}\label{(26)}
\begin{split}
&\Delta \Phi_p =2\pi\times\\& \left(\frac{\sqrt{-\frac{\left(r^{4} \Lambda +3 q^{2} {\mathrm e}^{-\gamma}-3 M r \right) r^{2}}{6 q^{2} {\mathrm e}^{-\gamma}+3 r^{2}-9 M r}}\, \sqrt{6}}{r^{2} \sqrt{\frac{\zeta}{\left(q^{2} {\mathrm e}^{-\gamma}+\frac{r \left(r -3 M \right)}{2}\right) r^{4}}}}-1\right), \\
\end{split} 
\end{equation}
\begin{equation}\label{(27)}
\begin{split}
&A =\frac{1}{3 {\mathrm e}^{-\gamma} q^{2} r^{2}+r^{3} \left(\Lambda  r^{3}-3 M \right)}\\&[12 q^{2} \left(\Lambda  r^{3}-\frac{9 M}{4}\right) r {\mathrm e}^{-\gamma}+4 \Lambda  r^{6}-\\ &15 \Lambda  r^{5} M +12 {\mathrm e}^{-2 \gamma} q^{4}-3 r^{3} M +18 r^{2} M^{2}].\\
\end{split}
\end{equation}
Given the chosen values for mass, $\gamma$, and $\Lambda$, the charge limit of our extremal black hole appears at q=2.48019052524.
\begin{figure}
\begin{center}
\subfigure[]{
\includegraphics[width=\linewidth]{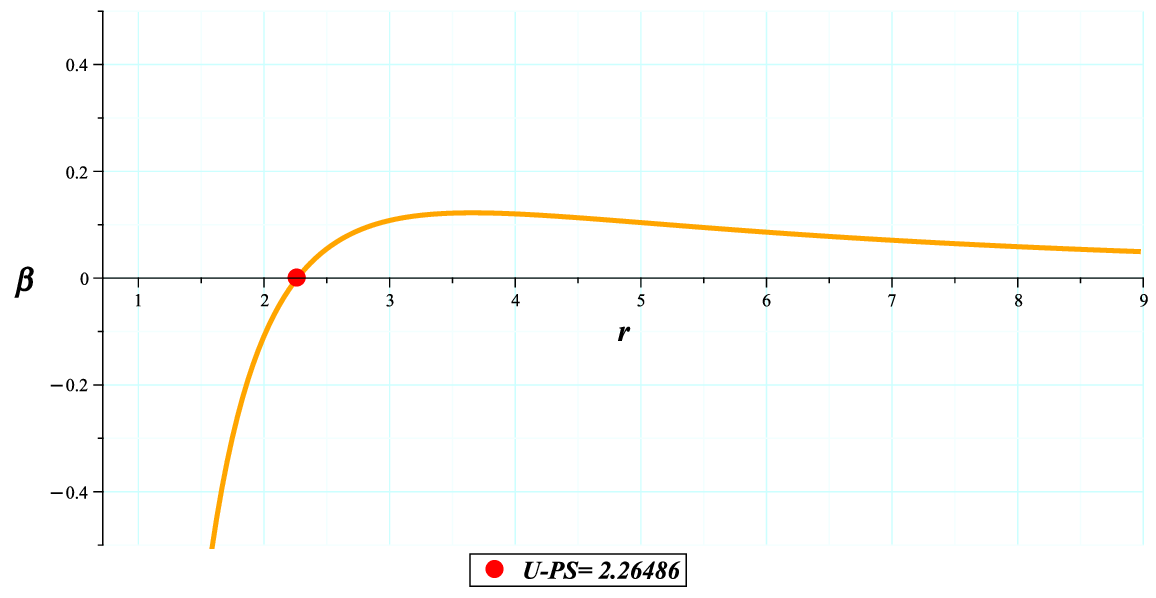}
\label{6a}}
\subfigure[]{
\includegraphics[width=\linewidth]{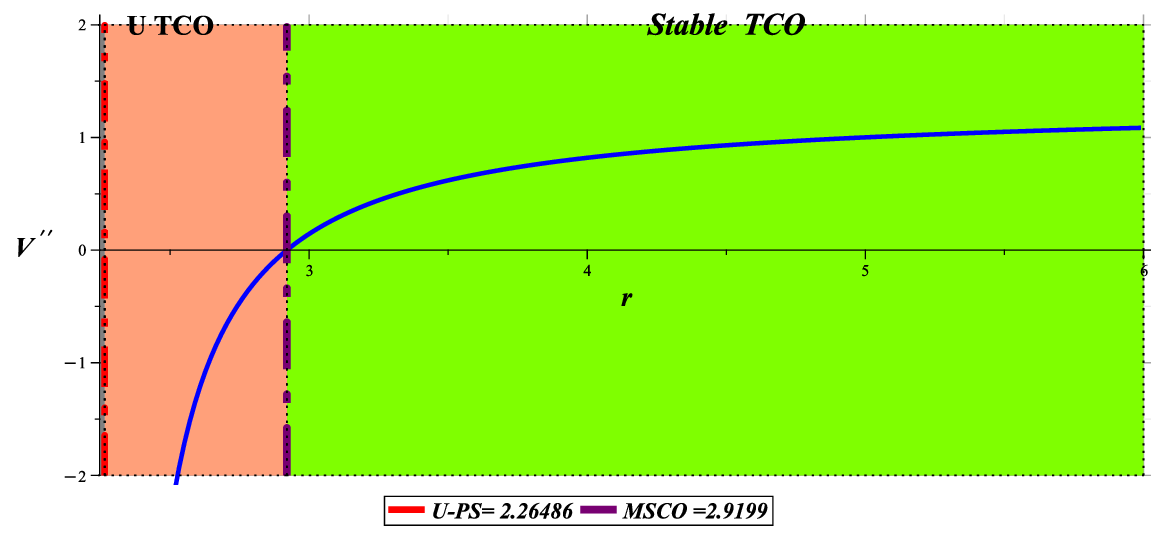}
\label{6b}}
\caption{\small{ (6a): $\beta$ diagram  (6b): MSCO localization and space classification with respect to $ M = 1, \gamma = 2,q = 2.48019,\Lambda = -1 $ for ModMax Black Hole}}
\label{6}
\end{center}
\end{figure}
From  Fig. (\ref{6}), it can be easily guessed that the ModMax model in its AdS form also follows the black hole behavior pattern well in the extremal limit. So adding the AdS radius ($\Lambda= 3/l^2 $) will not have any behavioral effect? To answer this question, it is better to look at the behavior of the "A" and $\sigma$ functions.
\begin{figure}
\begin{center}
\subfigure[]{
\includegraphics[width=\linewidth]{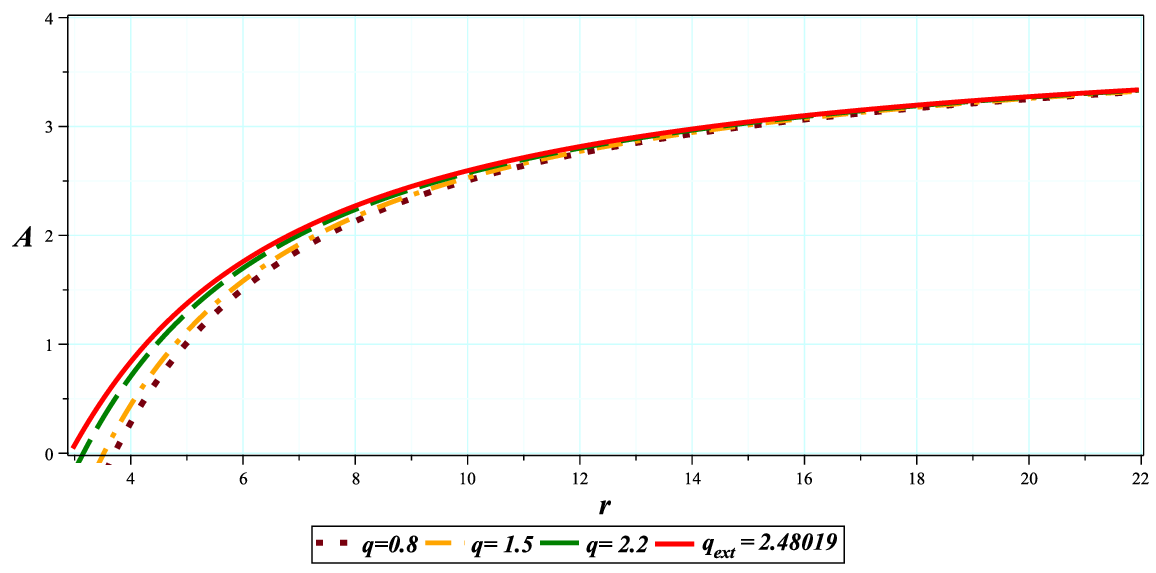}
\label{7a}}
\subfigure[]{
\includegraphics[width=\linewidth]{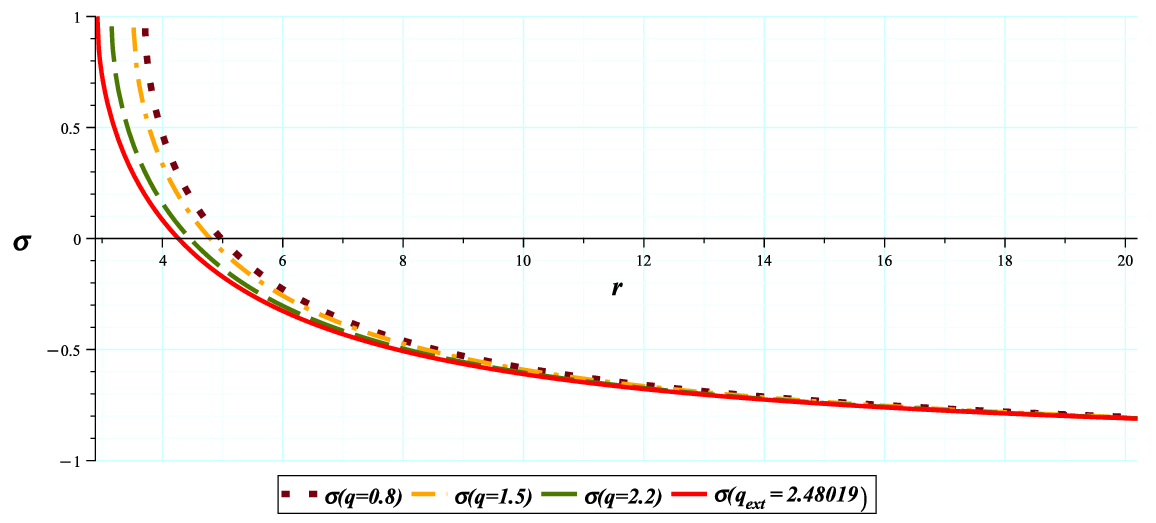}
\label{7b}}
\caption{\small{ (7a): "$A$" diagram  (7b): The $\sigma$ diagram with respect to $ M = 1, \gamma = 2 ,\Lambda = -1 $ with different q's for AdS ModMax Black Hole}}
\label{7}
\end{center}
\end{figure}
The key observation in Fig. (\ref{7}) is that the introduction of the AdS radius enables the parameter "A" to exceed unity as the distance from the horizon increases, thereby allowing "$\sigma$" to assume negative values. Consequently, the spacetime beyond the horizon bifurcates into two distinct regions: an inner region near the horizon where only prograde orbital motion is permitted, and an outer region sufficiently far from the horizon where only retrograde motion dominates, Fig. (\ref{8}). This transition underscores the role of spacetime curvature in shaping orbital dynamics.
\section{The Periapsis Shift in the presence of Aschenbach-Like Effect}
\begin{center}
\textbf{4D Born-Infeld Massive Gravity with a Non-Abelian Hair}
\end{center}
As discussed in the Introduction, some black-hole solutions admit a stable photon sphere (S-PS) outside the event horizon while preserving a regular black-hole geometry. The presence of an S-PS reflects the existence of a gravitational potential minimum outside the event horizon and gives rise to a non-monotonic orbital velocity profile, known as the Aschenbach-like effect \cite{27}. Black-hole solutions in massive gravity provide a natural setting in which this behaviour may occur \cite{28}. Motivated by these observations, we investigate AdS black holes in Born--Infeld massive gravity with non-Abelian hair\cite{46}.\\Our analysis addresses two questions:\\
(1) Does this spacetime admit a stable photon sphere outside the event horizon?\\
(2) If so, how does its presence modify the periapsis shift?\\
For brevity, the metric function and the corresponding components of the vector field $\Psi$ used in the photon-sphere analysis are presented in Appendix A, while only the quantities directly relevant to the discussion are given in the main text.
\begin{figure}
\begin{center}
\subfigure[]{
\includegraphics[width=\linewidth]{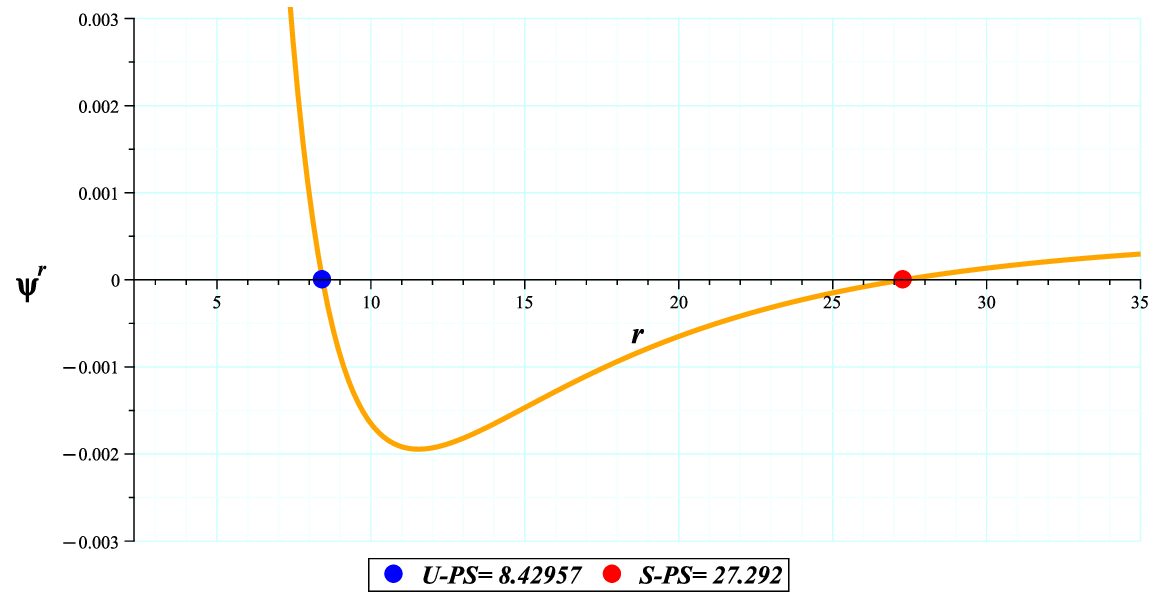}
\label{9a}}
\subfigure[]{
\includegraphics[width=\linewidth]{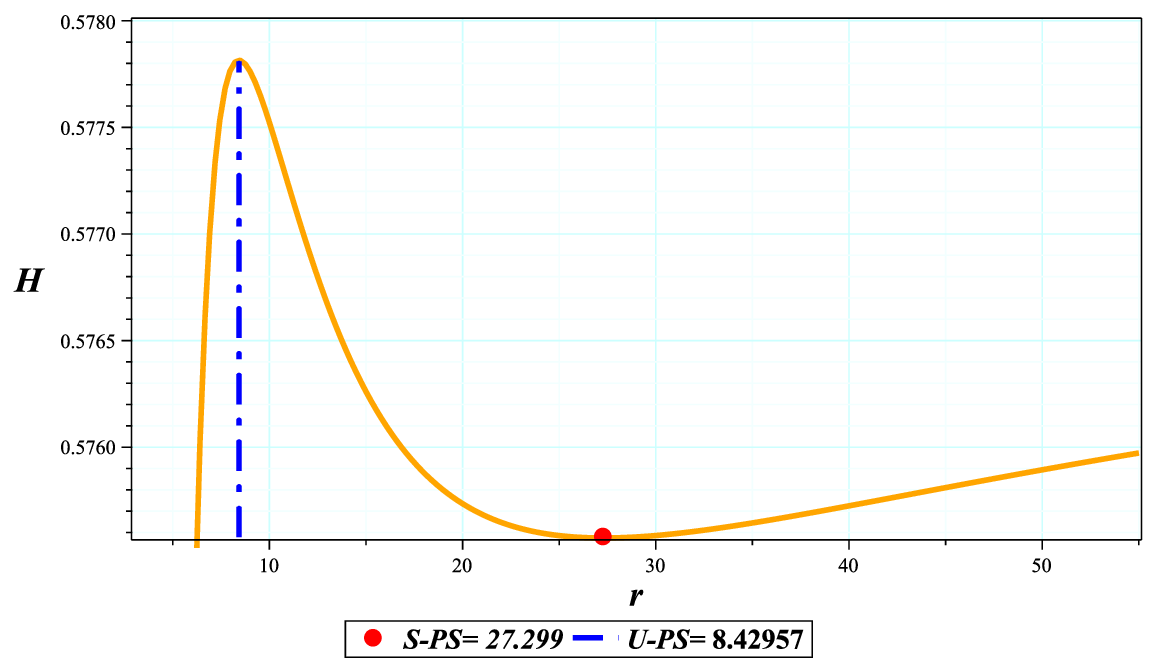}
\label{9b}}
\caption{\small{ (9a):  The U-PS and S-PS  with respect to $ M = 10, \zeta = 1, \nu = 1, \Lambda = -1, q = 1, C = 5, c_{1} = -5, c_{2} = 5, m_{g} = 0.1 $ , (9b): The topological potential H(r) for AdS black holes in Born-Infeld massive gravity }}
\label{9}
\end{center}
\end{figure}
Unlike the case shown in Fig.~(\ref{2}), the spacetime now admits two photon spheres outside the event horizon. As illustrated by the effective potential $H$ in Fig.~(\ref{9}), they correspond to a local maximum (U-PS) and a local minimum (S-PS), respectively. The existence of an external stable photon sphere indicates that this massive-gravity solution satisfies the geometric condition required for the emergence of the Aschenbach-like effect.
Now, according to Eq. (\ref{(5)}) and Eq. (\ref{(6)}), we plot the $\beta$ function and the second derivative of the potential for this model.
\begin{figure}
\begin{center}
\subfigure[]{
\includegraphics[width=\linewidth]{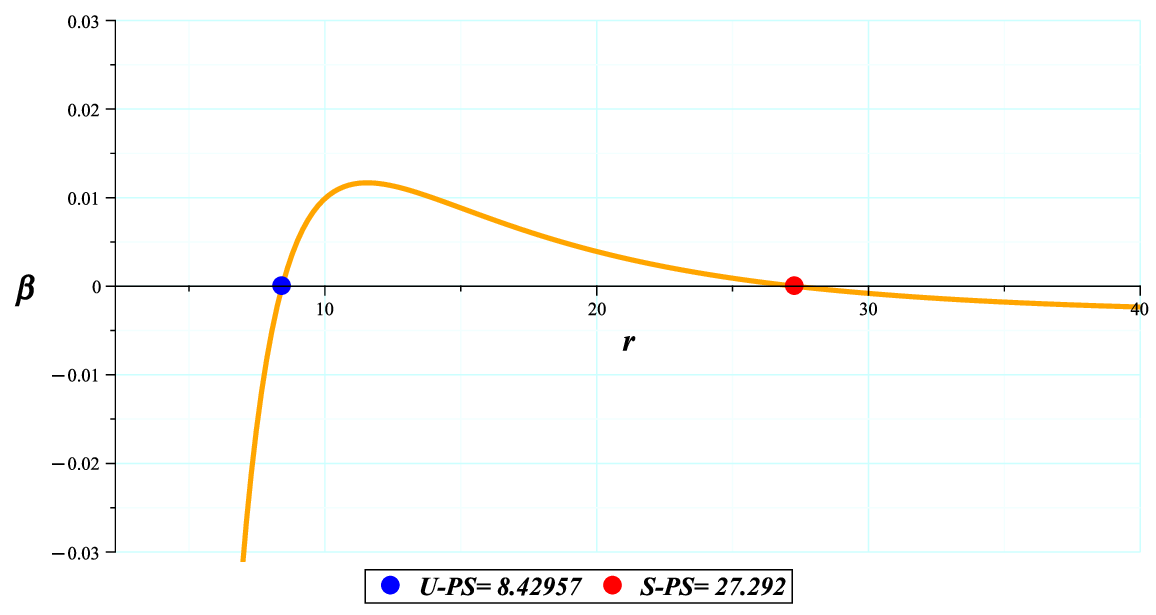}
\label{10a}}
\subfigure[]{
\includegraphics[width=\linewidth]{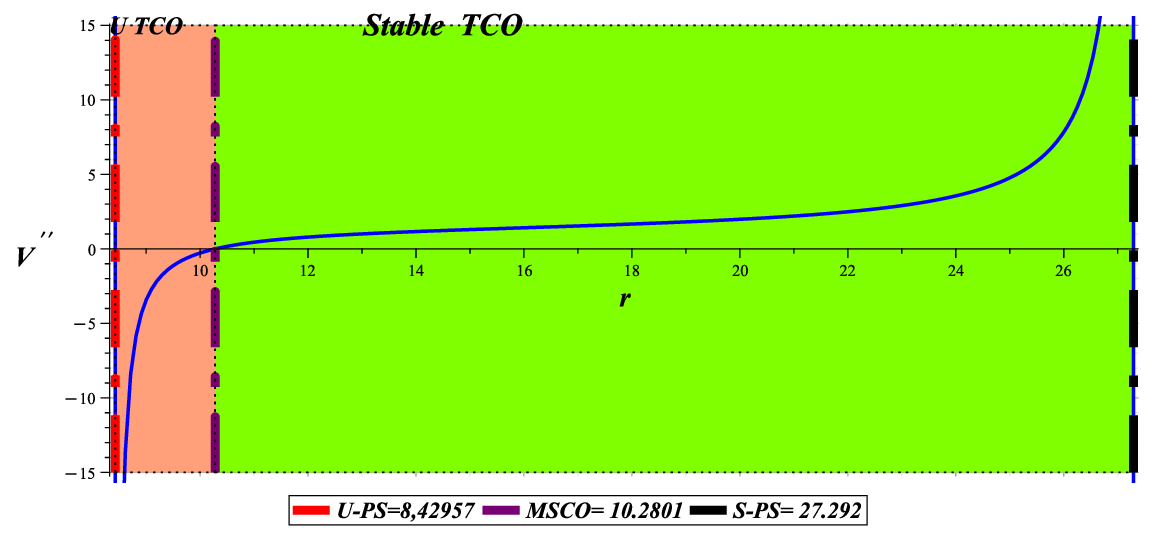}
\label{10b}}
\caption{\small{ (10a): $\beta$ diagram  (10b): MSCO localization and space classification with respect to $ M = 10, \zeta = 1, \nu = 1, \Lambda = -1, q = 1, C = 5, c_{1} = -5, c_{2} = 5, m_{g} = 0.1 $ for AdS black holes in Born-Infeld massive gravity}}
\label{10}
\end{center}
\end{figure}
Comparison of the $\beta$ function and the derivative of the effective potential in Fig.~(\ref{10}) with the previous cases highlights several qualitative differences. Unlike the previous models, where TCOs  extended from the unstable photon sphere through the MSCO  to arbitrarily large radii, the present solution admits TCOs only within the region bounded by the two photon spheres. Outside this interval, $\beta<0$, indicating that no physical TCOs exist. A further difference is evident from Fig.~(\ref{10b}) when compared with previous cases. In the previous models, the derivative of the effective potential approaches its asymptotic behaviour smoothly at large radii. By contrast, the present model exhibits a rapid increase in the second derivative of the potential near the stable photon sphere. This behaviour indicates a strong local variation in the effective potential near the stable photon sphere, which is expected to influence the corresponding orbital dynamics. These features distinguish the present solution from the previous cases and suggest that the neighbourhood of the stable photon sphere possesses qualitatively different orbital properties.
\begin{figure}
\begin{center}
\subfigure[]{
\includegraphics[width=\linewidth]{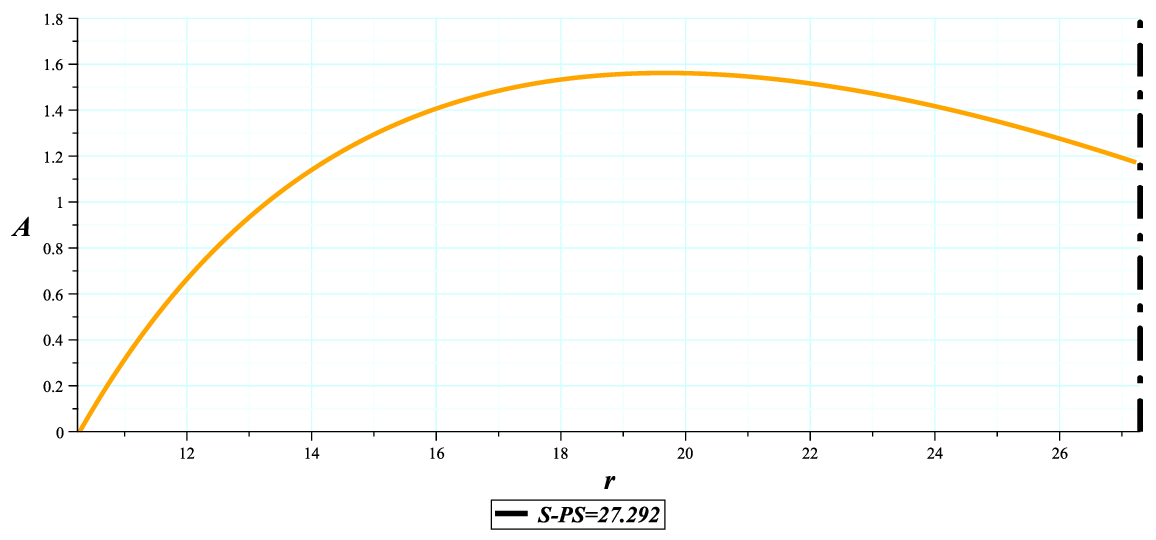}
\label{11a}}
\subfigure[]{
\includegraphics[width=\linewidth]{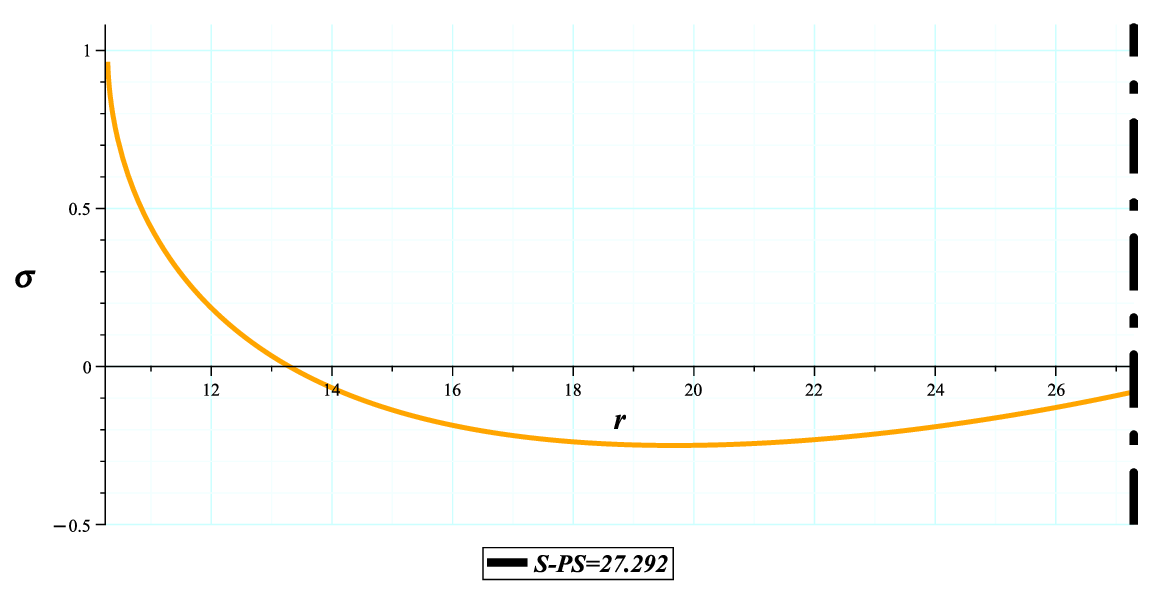}
\label{11b}}
\subfigure[]{
\includegraphics[width=\linewidth]{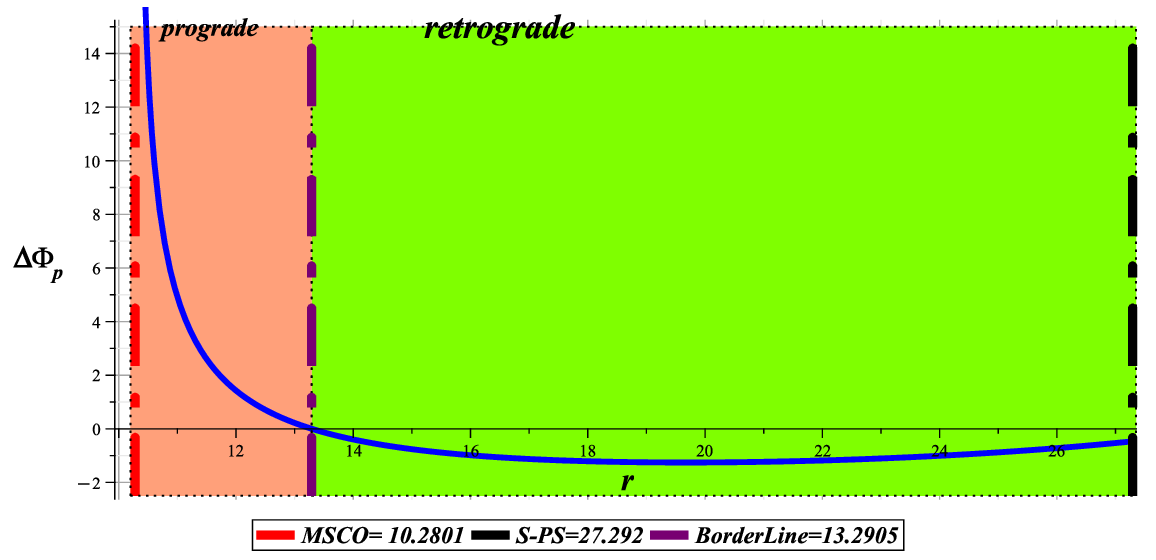}
\label{11c}}
\subfigure[]{
\includegraphics[width=\linewidth]{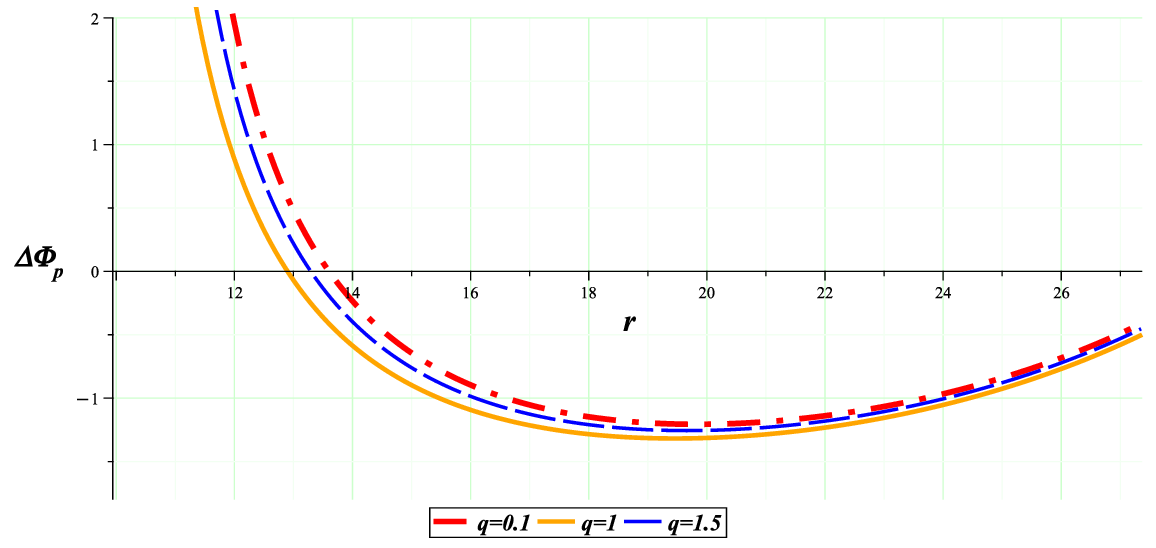}
\label{11d}}
\caption{\small{ (11a): "$A$" diagram  (11b): The $\sigma$ diagram and  (11c): $\Delta \Phi_{p}$ diagram with respect to $ M = 10, \zeta = 1, \nu = 1, \Lambda = -1, q = 1, C = 5, c_{1} = -5, c_{2} = 5, m_{g} = 0.1 $  (11d): $\Delta \Phi_{p}$ diagram  with different q's for AdS black holes in Born-Infeld massive gravity}}
\label{11}
\end{center}
\end{figure}
The behavior shown in Figure ~(\ref{11}) has qualitative similarities and differences with the previous configuration. As in the ModMax AdS black hole, the parameter $A$ exceeds unity over part of the allowed radial range (Fig.~(\ref{11a})), leading to the coexistence of prograde and retrograde orbital precession in different radial intervals (Fig.~(\ref{11c})). The behaviour nevertheless differs qualitatively from that observed in the previous models. As discussed in connection with Fig.~(\ref{10b}), the presence of the stable photon sphere is accompanied by a pronounced modification of the effective potential in its vicinity. This difference is also reflected in the behaviour of the parameters $A$ and $\sigma$. By comparison with Fig.~(\ref{7}):\\
-Unlike the AdS ModMax case, where $A$ approaches its asymptotic behaviour monotonically, the Born--Infeld solution exhibits a local maximum, followed by a decrease as the stable photon sphere is approached (Fig.~(\ref{11a})).\\
-A similar qualitative change is observed for $\sigma$. Rather than approaching its asymptotic value monotonically, $\sigma$ reaches a local minimum before increasing again in the neighbourhood of the stable photon sphere (Fig.~(\ref{11b}))\\
-For the parameter values considered here, the solution approaches a naked-singularity configuration near $q\simeq3$. Consequently, for the black-hole configurations studied in this work, the physically allowed region of $\Delta\Phi_p$ consists of two disconnected intervals: a prograde regime close to the event horizon and a retrograde regime at larger radii (Fig.~(\ref{11d})).\\
These qualitative differences are consistent with the appearance of a gravitational potential minimum outside the event horizon. The existence of this additional extremum modifies the local orbital structure and is accompanied by non-monotonic behaviour in the periapsis shift and is consistent with the Aschenbach-like orbital structure discussed previously.
\section{The Periapsis Shift in the presence of both Aschenbach-Like Effect and Extremality}
\begin{figure}
\begin{center}
\subfigure[]{
\includegraphics[width=\linewidth]{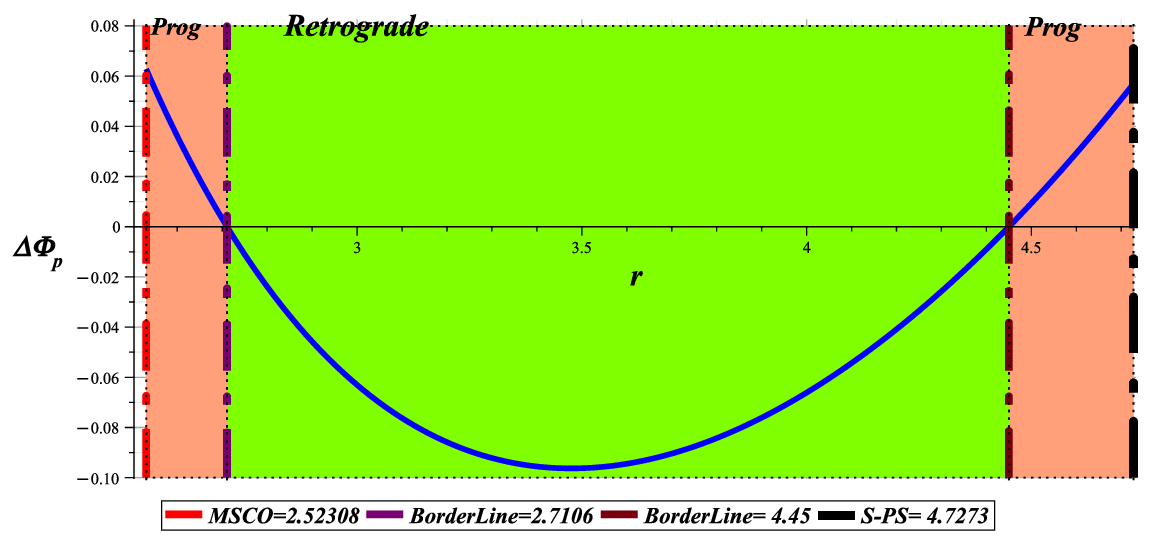}
\label{14a}}
\caption{\small{  $\Delta \Phi_{p}$ diagram with respect to $ M = 2, \gamma = 2,q=2.08817, \Lambda = -0.5, C = 0.4, c_{1} = -12, c_{2} = 25, m_{g} = 0.5 $ for  ModMax-dRGT-like massive gravity}}
\label{14}
\end{center}
\end{figure}

We now consider black-hole configurations in which extremality and the Aschenbach-like effect coexist, allowing us to investigate their combined influence on orbital dynamics. The spacetime must simultaneously satisfy two conditions. First, it must admit a well-defined extremal black-hole configuration with a regular event horizon. Second, it must support the geometric structure required for the Aschenbach-like effect, namely the existence of a stable photon sphere outside the horizon. Under these conditions, we investigate how the periapsis shift is modified when both properties are present. To this end, we revisit the massive-gravity solution previously identified as exhibiting the Aschenbach-like effect \cite{28} and extend the analysis to its extremal limit.
\begin{center}
\textbf{4D ModMax-dRGT-like massive gravity}
\end{center}
Our calculations show that the superextremal charge limit for this model \cite{47}, given the selected parameter values, appears at  $q=2.0881742$ (For brevity, the metric function and its diagram are presented in Appendix A). According to Eq. (\ref{(5)}) and Eq. (\ref{(6)}), the $\beta$ function and the second derivative of the potential for this model will be as:
\begin{equation}\label{(32)}
\begin{split}
&\beta =\\ &(24 q^{2} {\mathrm e}^{-\gamma}+3 C \,r^{3} c_{1} m_{g}^{2}+\left(12 C^{2} c_{2} m_{g}^{2}+12\right) r^{2}-18 M r)\times\\ &(\frac{1}{6 q^{2} {\mathrm e}^{-\gamma}-2 r^{4} \Lambda +3 C \,r^{3} c_{1} m_{g}^{2}+\left(6 C^{2} c_{2} m_{g}^{2}+6\right) r^{2}-6 M r}),\\
\end{split}
\end{equation}
\begin{equation*}\label{(0)}
\begin{split}
&\alpha =\\ &18 C^{3} r^{5} c_{1} c_{2} m_{g}^{4}+3 C^{2} r^{6} c_{1}^{2} m_{g}^{4}-32 C^{2} \Lambda  r^{6} c_{2} m_{g}^{2}-6 C \Lambda  r^{7} c_{1} m_{g}^{2}\\ & +66 C {\mathrm e}^{-\gamma} q^{2} r^{3} c_{1} m_{g}^{2}+12 C^{2} r^{3} c_{2} M m_{g}^{2}+18 C \,r^{5} c_{1} m_{g}^{2},\\
\end{split}
\end{equation*}
\begin{equation*}\label{(0)}
\begin{split}
&\varrho =36 C \,r^{4} c_{1} M m_{g}^{2}-96 \Lambda  {\mathrm e}^{-\gamma} q^{2} r^{4}-32 \Lambda  r^{6}+\\ &60 \Lambda  r^{5} M -96 {\mathrm e}^{-2 \gamma} q^{4}+108 {\mathrm e}^{-\gamma} q^{2} r M +12 r^{3} M -36 r^{2} M^{2},\\
\end{split}
\end{equation*}

\begin{equation}\label{(33)}
\begin{split}
\mathbb{V}''=\frac{\alpha -\varrho}{6 \left(4 C^{2} r^{2} c_{2} m_{g}^{2}+C \,r^{3} c_{1} m_{g}^{2}+8 q^{2} {\mathrm e}^{-\gamma}+4 r^{2}-6 M r \right) r^{4}}.
\end{split}
\end{equation}
\begin{figure}
\begin{center}
\subfigure[]{
\includegraphics[width=\linewidth]{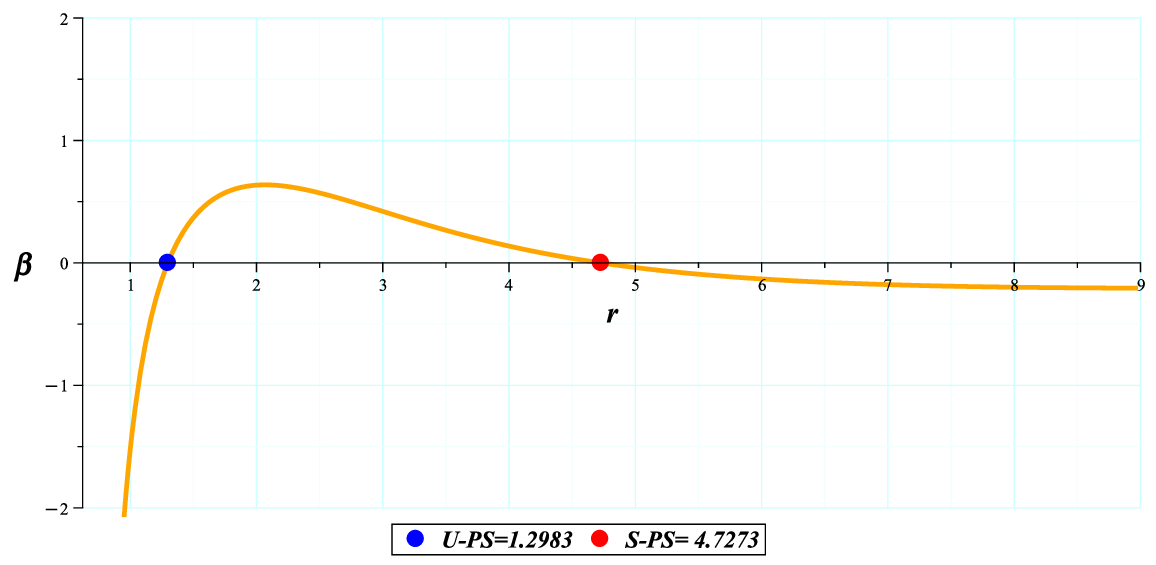}
\label{12a}}
\subfigure[]{
\includegraphics[width=\linewidth]{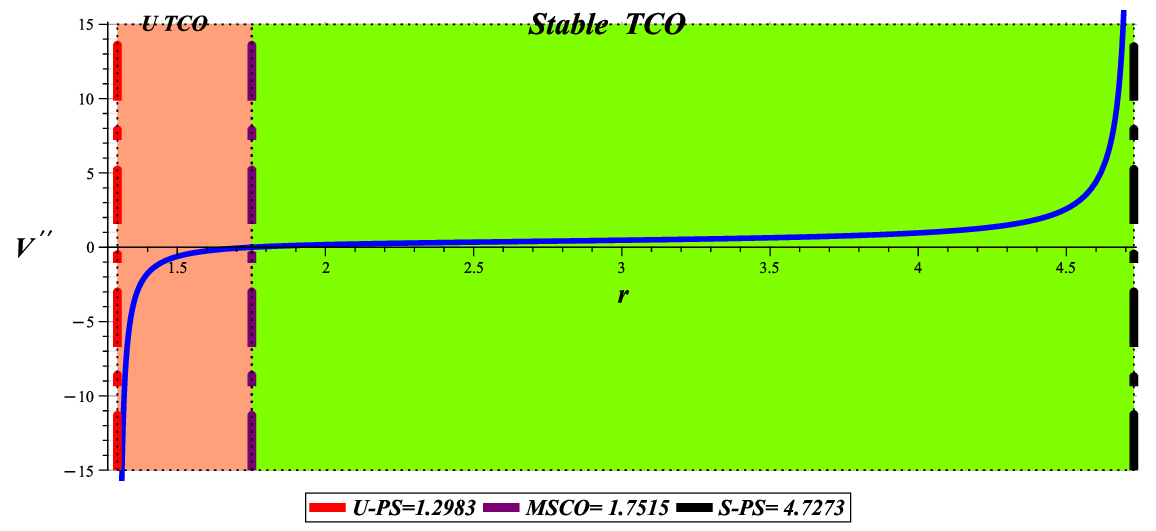}
\label{12b}}
\caption{\small{ (12a): $\beta$ diagram  (12b): MSCO localization and space classification with respect to $ M = 2,  \gamma = 2, \Lambda = -0.5, q = 2.0881742, C = 0.4, c_{1} = -12, c_{2} = 25, m_{g} = 0.5 $ for AdS black holes in ModMax-dRGT-like massive gravity}}
\label{12}
\end{center}
\end{figure}
Comparing Fig. (\ref{12}) with Fig. (\ref{10}) shows that, although the model is extremal here, it accurately exhibits black hole behavior. That is, in addition to the appearance of the photon spheres beyond the event horizon Fig. (\ref{12a}), it exhibits the geodesic behavior like the previous Born-Infeld black hole. Here too, the change and rise of the effective potential energy near the S-PS can be observed,Fig. (\ref{12b}).
\begin{figure}
\begin{center}
\subfigure[]{
\includegraphics[width=\linewidth]{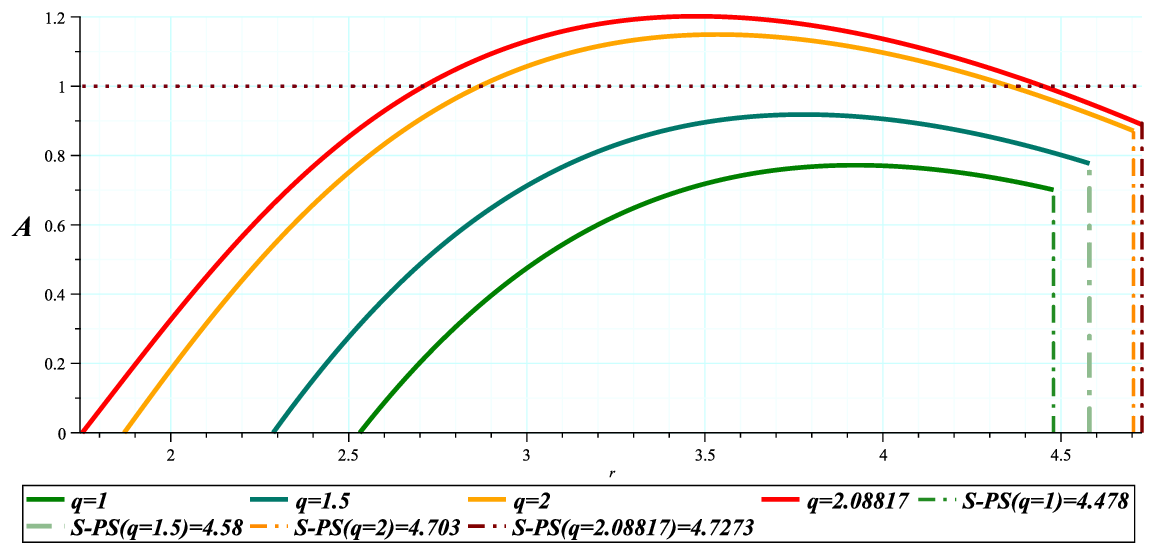}
\label{13a}}
\subfigure[]{
\includegraphics[width=\linewidth]{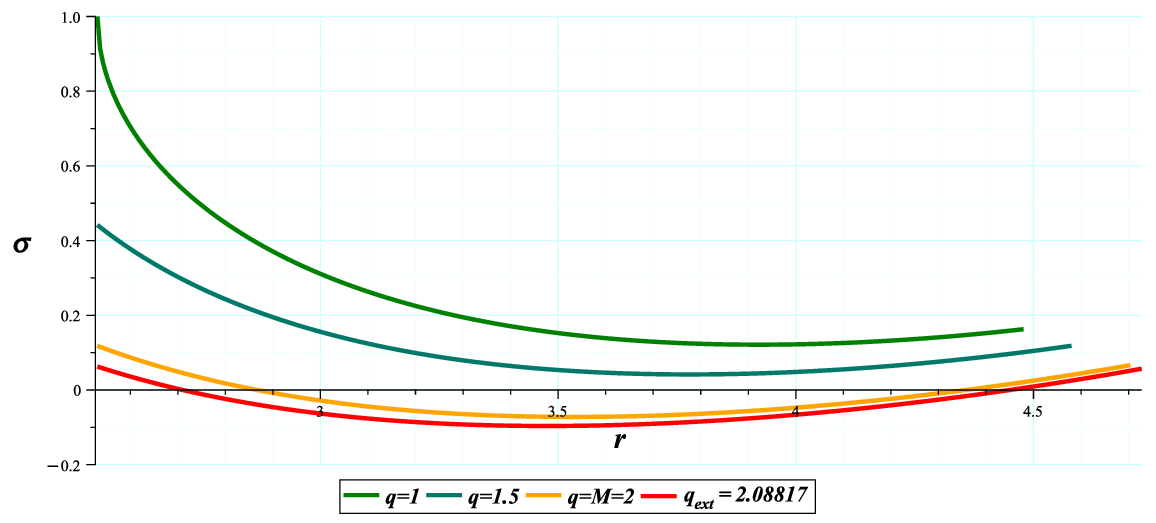}
\label{13b}}
\subfigure[]{
\includegraphics[width=\linewidth]{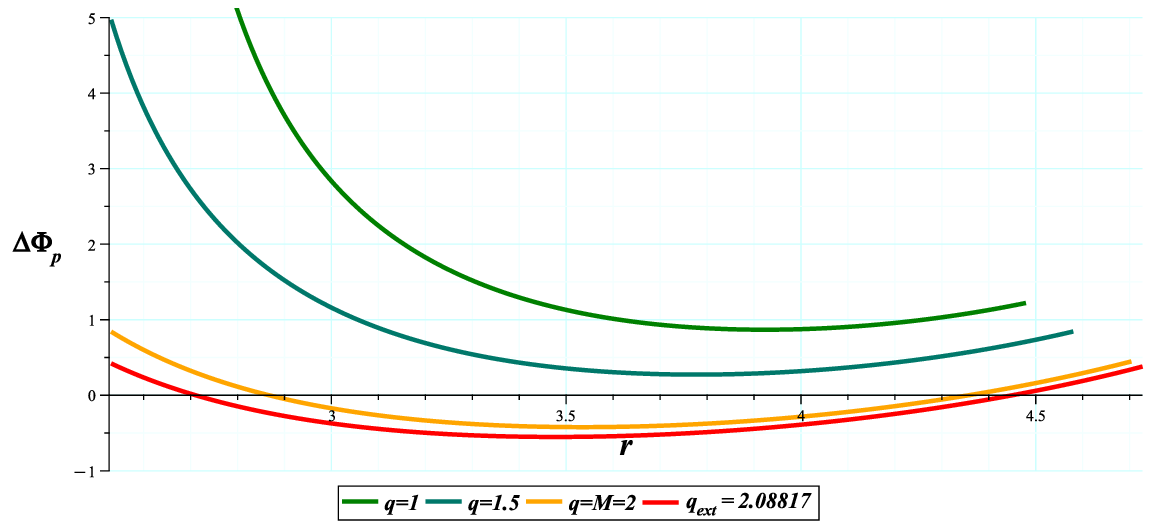}
\label{13c}}
\caption{\small{ (13a): "$A$" diagram  (13b): The $\sigma$ diagram and  (13c): $\Delta \Phi_{p}$ diagram with respect to $ M = 2, \gamma = 2, \Lambda = -0.5, C = 0.4, c_{1} = -12, c_{2} = 25, m_{g} = 0.5 $ with different $q$ for  AdS black holes in  ModMax-dRGT-like massive gravity}}
\label{13}
\end{center}
\end{figure}
Figure~(\ref{13}) shows the behaviour of $A$, $\sigma$, and $\Delta\Phi_p$ for subextremal, extremal, and superextremal configurations obtained from Eq.~(\ref{(14)}). Figure~(\ref{13a}) shows that $A$ exceeds unity over part of the allowed radial range, implying the coexistence of prograde and retrograde orbital precession.
Compared with the ModMax and Born-Infeld solutions  the present model exhibits a qualitatively different orbital structure.\\
In the subextremal regime, the periapsis shift remains entirely prograde. As the extremal limit is approached, a retrograde region develops, followed by the reappearance of prograde motion at larger radii. The reappearance of prograde motion coincides with the vicinity of the stable photon sphere, where the effective potential develops an external minimum.\\
Consequently, the spacetime is divided into three distinct orbital regions: an inner prograde region close to the event horizon, an intermediate retrograde region, and an outer prograde region beyond the stable photon sphere (Fig.~(\ref{14})).
The recovery of prograde motion beyond the retrograde region is associated with the appearance of the external potential minimum near the stable photon sphere, indicating a qualitative modification of the orbital structure.
A similar tendency is present in the Born-Infeld solution ; however, in the present model the transition between the different orbital regimes is considerably more pronounced. The coexistence of extremality and the Aschenbach-like effect therefore gives rise to a three-region orbital structure, a feature not observed in the other black-hole models considered in this work.
\section{Conclusion}
This work has explored the behaviour of the periapsis shift in several charged black-hole geometries, with particular emphasis on two complementary aspects: the approach to extremality and the presence of the Aschenbach-like effect. Rather than treating these phenomena independently, we investigated how they jointly influence the orbital dynamics of test particles in static black-hole spacetimes.  Our analysis shows that the periapsis shift remains a well-defined and informative dynamical quantity as the extremal limit is approached. In the models considered here, extremality does not eliminate the characteristic orbital structures associated with black holes. Instead, the evolution of the periapsis shift continuously reflects the underlying spacetime geometry, providing additional evidence that the orbital dynamics remain physically meaningful even near the extremal configuration.  For black-hole solutions admitting a stable photon sphere outside the event horizon, the orbital behaviour undergoes a qualitative change. In these cases, the emergence of an external minimum in the effective potential, previously associated with the Aschenbach-like effect, modifies both the angular-velocity profile and the periapsis shift. This demonstrates that the influence of the stable photon sphere extends beyond purely geometric considerations and leaves a direct imprint on observable orbital dynamics.  The most distinctive result of this study is obtained when extremality and the Aschenbach-like effect coexist. Under these conditions, the periapsis shift exhibits a three-region orbital structure consisting of an inner prograde region, an intermediate retrograde region, and a second prograde region at larger radii. Within the black-hole models investigated in this work, such behaviour is absent from the other configurations considered, indicating that the combined presence of extremality and an external stable photon sphere gives rise to a qualitatively different orbital evolution.  These results also illustrate the advantage of the periapsis shift as a dynamical diagnostic of black-hole spacetimes. While the existence of photon spheres primarily characterizes the geometric structure of the spacetime, the periapsis shift simultaneously encodes both geometric and dynamical information, making it a more sensitive probe of changes in the effective potential and the associated orbital motion.  Although the present analysis is restricted to specific classes of static charged black holes, the framework developed here can be applied to a broader range of gravitational theories. Extending this investigation to rotating black holes or to alternative modified-gravity models may help determine whether the three-region orbital structure identified here represents a generic consequence of the interplay between extremality and non-monotonic orbital dynamics or is instead characteristic of a more restricted class of black-hole solutions.

\section{Appendix A}
\subsection*{4D AdS ModMax Black Hole}
The spherically-symmetric static metric line element of such a spacetime is given by :
\begin{equation}\label{(23)}
e^{\nu(r)}=e^{-\lambda(r)}=1-\frac{2 M}{r}+\frac{q^{2} {\mathrm e}^{-\gamma}}{r^{2}}-\frac{r^{2} \Lambda}{3},
\end{equation}
where $\Lambda$ is a cosmological constant.
\begin{figure}
\begin{center}
\subfigure[]{
\includegraphics[width=\linewidth]{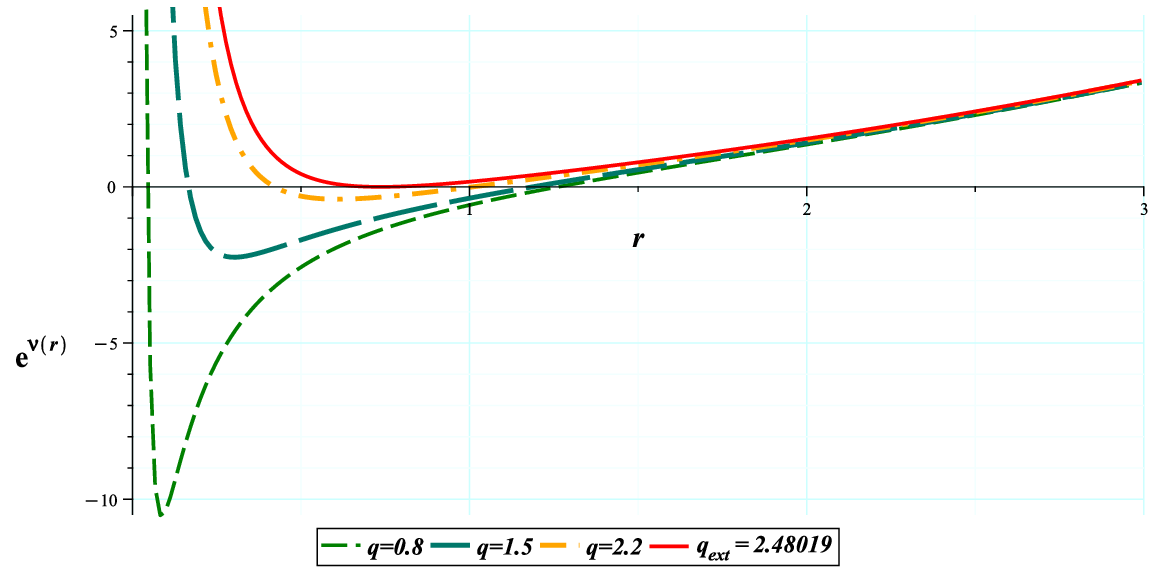}
\label{15a}}
\subfigure[]{
\includegraphics[width=\linewidth]{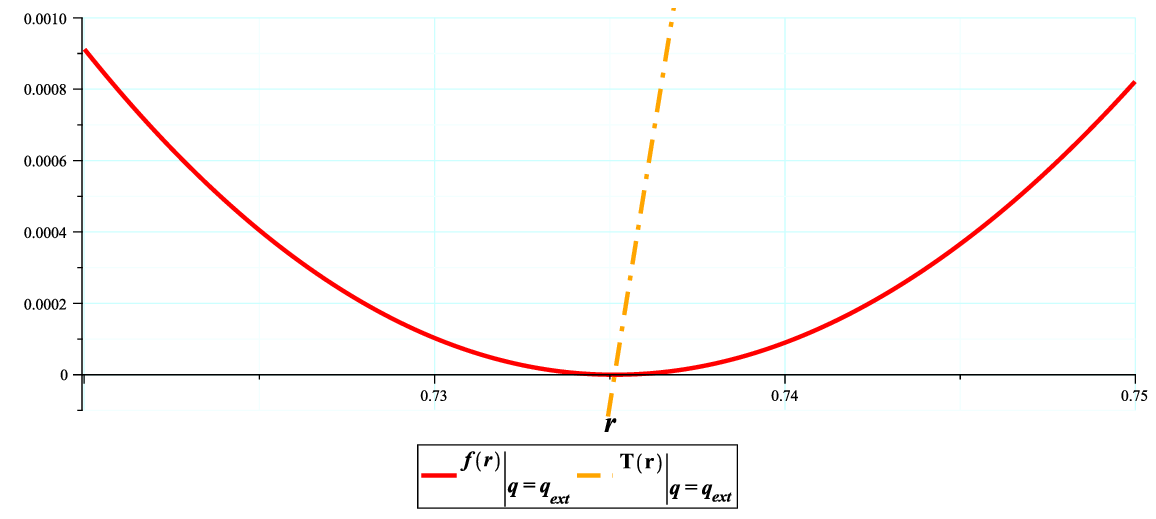}
\label{15b}}
\subfigure[]{
\includegraphics[width=\linewidth]{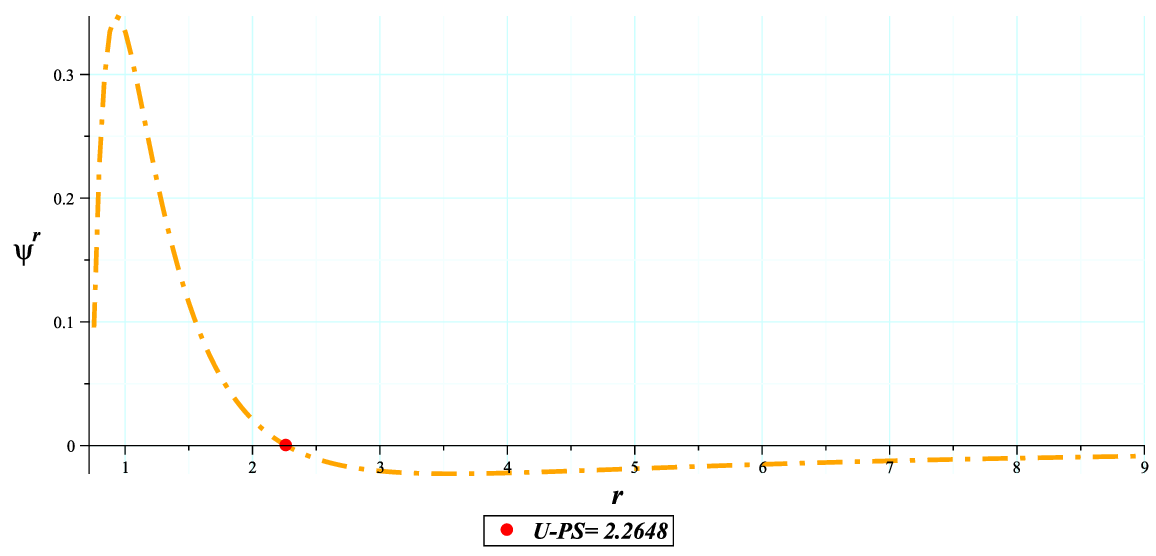}
\label{15c}}
\subfigure[]{
\includegraphics[width=\linewidth]{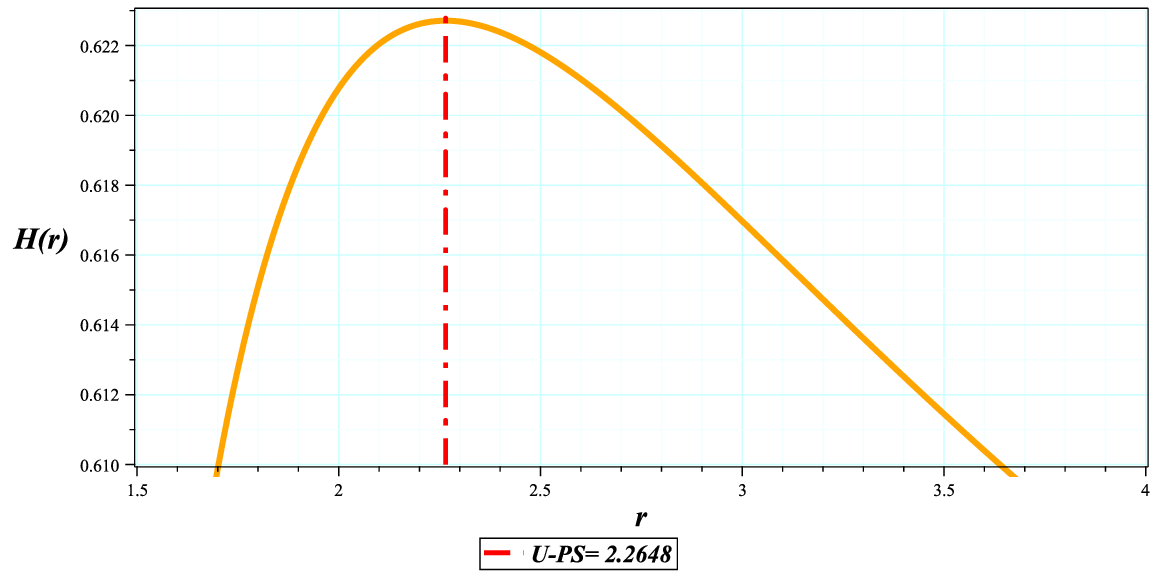}
\label{15d}}
\caption{\small{ (15a): Metric function with $ M = 1, \Lambda = -1, \gamma = 2$ and different q (15b): Metric function VS temperature at the extreme limit  (15c): The unstable photon spheres(U-PS)  with respect to $ M = 1,\Lambda = -1, \gamma = 2,q = 2.48019 $ , (15d): the topological potential H(r) for AdS ModMax Black Hole }}
\label{15}
\end{center}
\end{figure}
In Fig. (\ref{15}), we show the metric function for different charges, the behavior of the temperature function and the metric function in the extremal limit, the representation of the geometric locus of U-PS, and finally the maximum of this photon sphere from the perspective of the effective potential.

\subsection*{Long Equations for Born-Infeld Massive Gravity}
The metric function for AdS black holes in Born-Infeld massive gravity with non-abelian hair \cite{46} is:
\begin{equation}\label{(A1)}
\begin{split}
&e^{\nu(r)}=e^{-\lambda(r)}=\\&1-\frac{M}{r}-\frac{r^{2} \Lambda}{3}+\frac{\nu^{2}}{r^{2}}+\left(\frac{r c_{1}}{2}+C c_{2}\right) C m_{g}^{2}+\\&\frac{2 \left(1-\mathrm{hypergeom}\! \left(\left[-\frac{3}{4},-\frac{1}{2}\right],\left[\frac{1}{4}\right],-\frac{q^{2}}{r^{4} \zeta^{2}}\right)\right) r^{2} \zeta^{2}}{3},\\
\end{split}
\end{equation}
where $\nu$ is the magnetic parameter, $q$ electric charge, $\zeta$ nonlinearity parameter, $M$ black hole mass, $\Lambda$ cosmological constant, $m_g$ graviton mass, and $c_i$ free constants.

The topological potential $H$ for this model is:
\begin{equation*}
\begin{split}
\varsigma= 9-\frac{9 M}{r}-3 r^{2} \Lambda +\frac{9 \nu^{2}}{r^{2}}+9 \left(\frac{r c_{1}}{2}+C c_{2}\right) C m_{g}^{2},
\end{split}
\end{equation*}
\begin{equation}\label{(A2)}
\begin{split}
&H =\\ & \frac{\sqrt{\varsigma+6 \left(1-\mathrm{hypergeom}\! \left(\left[-\frac{3}{4},-\frac{1}{2}\right],\left[\frac{1}{4}\right],-\frac{q^{2}}{r^{4} \zeta^{2}}\right)\right) r^{2} \zeta^{2}}}{3 \sin \! \left(\theta \right) r},\\
\end{split}
\end{equation}
and the radial component of the vector field $\Psi$ is:
\begin{equation*}
\iota=-\frac{\sqrt{2}\, \left(4 C^{2} r^{2} c_{2} m_{g}^{2}+C \,r^{3} c_{1} m_{g}^{2}-6 M r +8 \nu^{2}+4 r^{2}\right)}{4},
\end{equation*}
\begin{equation}\label{(A3)}
\Psi^{r}=\frac{\sqrt{2}\, \csc \! \left(\theta \right) \left(\iota-\frac{\pi  \mathrm{LegendreP}\left(-\frac{1}{4},-\frac{1}{4},\frac{r^{4} \zeta^{2}-q^{2}}{r^{4} \zeta^{2}+q^{2}}\right) q^{2}}{\left(-\frac{q^{2}}{r^{4} \zeta^{2}}\right)^{\frac{1}{8}} \left(\frac{r^{4} \zeta^{2}+q^{2}}{r^{4} \zeta^{2}}\right)^{\frac{1}{4}} \Gamma \left(\frac{3}{4}\right)}\right)}{2 r^{4}}.
\end{equation}
\subsection*{4D ModMax-dRGT-like massive gravity}
The spherically-symmetric static metric line element of such a spacetime is given by \cite{47}:
\begin{equation}\label{(31)}
e^{\nu(r)}=e^{-\lambda(r)}=1-\frac{M}{r}-\frac{r^{2} \Lambda}{3}+\frac{q^{2} {\mathrm e}^{-\gamma}}{r^{2}}+\left(\frac{r c_{1}}{2}+C c_{2}\right) C m_{g}^{2}
\end{equation}
where all the parameters in this equation have been previously introduced.
\begin{figure}
\begin{center}
\subfigure[]{
\includegraphics[width=\linewidth]{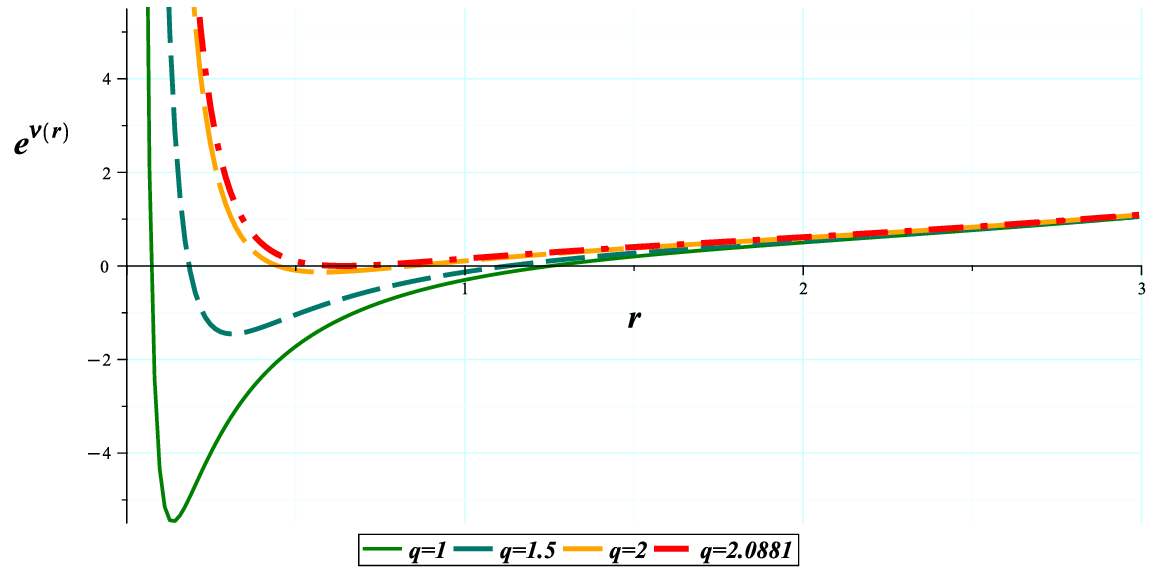}
 }
\caption{\small{ Metric function: With $ M = 2, \gamma = 2, \Lambda = -0.5, C = 0.4, c_{1} = -12, c_{2} = 25, m_{g} = 0.5 $ with different q's }}
\label{16}
\end{center}
\end{figure}
the structure clearly has a Cauchy and event horizon up to $q= 2.0881$,Fig. (\ref{16}).

\end{document}